\documentclass[preprint2]{aastex} 
\usepackage{graphicx, subfigure}
\usepackage{natbib}
\usepackage{times}
\usepackage{amsmath}
\usepackage{textcomp}
\usepackage{amssymb}
\shorttitle{High-$z$ BL Lacs}
\shortauthors{Kaur et al.}

\def\asec{\ifmmode^{\prime\prime}\else$^{\prime\prime}$\fi}
\def\farcs{\hbox{$.\!\!^{\prime\prime}$}}       
\def\degs{\ifmmode ^{\circ}\else$^{\circ}$\fi}

\begin{document}

\title{New High-$z$ Fermi BL Lacs with the Photometric Dropout Technique}
\author{A. Kaur\altaffilmark{1}, A. Rau\altaffilmark{2}, M. Ajello\altaffilmark{1},  J. Greiner\altaffilmark{2},  D. H. Hartmann\altaffilmark{1}, V. S. Paliya\altaffilmark{1},A. Dom\'{i}nguez\altaffilmark{3}, J. Bolmer\altaffilmark{2,4}, P. Schady\altaffilmark{2}}
\affil{\altaffilmark{1}Department of Physics and Astronomy, Clemson University, SC 29634-0978, U.S.A.}
     \email{akaur@g.clemson.edu}  
     \affil{\altaffilmark{2}Max-Planck-Institut f\"ur extraterrestrische Physik, Giessenbachstra{\ss}e 1, 85748 Garching, Germany}
\affil{\altaffilmark{3}Grupo de Altas Energ\'{i}as, Universidad Complutense, E-28040 Madrid, Spain}
\and
\affil{\altaffilmark{4}European Southern Observatory, Alonso de C\'{o}rdova 3107, Vitacura, Casilla 19001, Santiago 19, Chile}
\begin{abstract}
Determining redshifts for BL Lacertae (BL Lac) objects using the traditional spectroscopic method is challenging due to the absence of strong emission lines in their optical spectra. We employ the photometric dropout technique to determine redshifts for this class of blazars using the combined 13 broad-band filters from {\it Swift}-UVOT and the multi-channel imager GROND at the MPG 2.2 m telescope at ESO's La Silla Observatory. The wavelength range covered by these 13 filters extends from far ultraviolet to the near-Infrared. We report results on 40 new Fermi detected BL Lacs with the photometric redshifts determinations for 5 sources, with 3FGL J1918.2-4110 being the most distant in our sample at $z$=2.16. Reliable upper limits are provided for 20 sources in this sample. Using the highest energy photons for these Fermi-LAT sources, we evaluate the consistency with the Gamma-ray horizon due to the extragalactic background light. 
 \end{abstract}
\keywords{(galaxies:) BL Lacertae objects: general --- galaxies: active --- gamma rays: diffuse background}

\section{Introduction}
Blazars represent a class of active galactic nuclei (AGNs) with relativistic jets pointing along our line of sight \citep{Blandford1978}. Their spectral energy distribution (SED) exhibits two characteristic broad bumps, which are attributed to synchrotron emission at the lower energies (Infrared to X-ray) and synchrotron self Compton at the higher energies \citep[X-ray to $\gamma$-rays, e.g.][]{Maraschi1994}. On the basis of their optical spectroscopic characteristics, blazars can be further classified into two types : Flat Spectrum Radio Quasars (FSRQs), characterized by broad emission lines and BL Lacertae objects (BL Lacs), with no or at best weak emission lines (equivalent width $<$ 5 \AA, \citealt{Urry1995}). Furthermore, another classification scheme for blazars was introduced by \citet{Abdo2010} based on the location of their synchrotron peak frequency,  $\nu_{pk}^{sy}$. The authors subdivided these objects into three classes: high-synchrotron-peaked blazars (HSP) for $\nu_{pk}^{sy} > 10^{15}$ Hz, intermediate-synchrotron-peak (ISP) for $10^{14} \mbox{ Hz} < \nu_{pk}^{sy} < 10^{15}$ Hz and low-synchrotron-peak (LSP) for $\nu_{pk}^{sy} < 10^{14}$ Hz. Most of the FSRQs fall in the LSP category, but half the population of BL Lacs display peak synchrotron frequencies $ > 10^{15}$ Hz \citep{Ackermann2015}. These high values of $\nu_{pk}^{sy}$ imply the presence of relativistic multi-TeV electrons, therefore making BL Lacs very bright $\gamma$-ray sources with substantial emission above 10 GeV \citep{Ackermann2013a}.\\

Blazars play an important role in the study of the Extragalactic Background Light (EBL), which represents the integrated light from all the stars and other compact objects since the re-ionization epoch. The photons from blazars are attenuated by EBL photons through  production of electron-positron pairs, which imprint a characteristic signature in the spectra of these $\gamma$-ray sources \citep{Stecker1992}. This feature in blazar spectra enables us to constrain the EBL and its evolution with cosmic time \citep{Aharonian2006, Ackermann2012, Dominguez2013}. In the case of FSRQs, the presence of broad emission lines implies the presence of a disk, whose UV radiation field could attenuate $\gamma$-ray photons, thereby making it a challenging task to differentiate the attenuation signal from the EBL photons from the circum-nuclear one. On the other hand, absence of broad emission lines in BL Lacs in addition to the abundance of photons above 10 GeV render them the perfect class of blazars to explore the EBL \citep{Dominguez2015}. To enable such studies, redshift measurements for these sources are essential. In particular, high redshift BL Lacs are critical for probing the EBL, since the strength of the attenuation  increases with redshift. Moreover, high-$z$ blazars are crucial for testing the EBL evolution, which at present is poorly constrained. Estimating redshifts of BL Lacs in the traditional spectroscopic way is yet another challenge, because of the weakness, or even absence of lines \citep{Shaw2013}. \\
\citet{Rau2012} initiated a program to determine redshifts for BL Lac sources via a photometric technique. The underlying principle for this photometric redshift (photo-$z$) determination is based on the absorption of UV photons by neutral hydrogen along the line of sight, which absorbs photons bluewards of the Lyman limit. This leads to a dropout in the flux at the Lyman limit whose position in the SED can be used to determine the photo-$z$. This approach was applied by \citet{Rau2012} for a sample of 103 blazars.   
 \\
 The immediate aim of this study is to determine photo-$z$s of high-$z$ ($z~>$ 1.3, which is the lower limit measurable with this method) BL Lacs to increase the sample in the important high redshift regime. Only 19 BL Lacs with $z~ >$ 1.3 are known, of which 13 are reported in the third catalog of Fermi-detected AGNs (3LAC, \citet{Ackermann2015}). A total of 9 BL Lacs were reported in \citet{Rau2012}, of which three sources are provided among the 13 in the 3LAC catalog. Therefore, \citet{Rau2012} provided 6 new high-$z$ BL Lacs with the photometric technique. \\ The outline of this paper is as follows: Section 2 explains the details of observations. The data analysis procedure is explained in section 3. The resulting high-$z$ BL Lacs parameters are reported in section 4 and the interpretation of our results is presented in section 5. A flat $\Lambda$CDM cosmological model with H$_0$=71 km s$^{-1}$ Mpc$^{-1}$, $\Omega_m$=0.27 and $\Omega_{\Lambda}$ =0.73 was adopted for all the calculations in this work in order to be consistent with the work presented in \citet{Finke2013}.

\section{Observations}
\label{sec:obs}
\subsection{Sample Selection}
Our sample was selected from sources classified as BL Lacs in the third catalog of Fermi detected sources (3FGL, \citealt{Acero2015}) without a measured redshift. The selection procedure was based on two main criteria, one being that only sources with Declination $<$ 25$^{\circ}$ were chosen in order to accommodate the visibility from the ground based telescope, situated in Chile. The second selection criterion was the minimization of the Galactic foreground reddening, ensured by selecting objects away from the Galactic Plane ($|b|$ $> $ 10 $^{\circ}$). Here we report on the 40 sources observed so far whose basic properties are presented in Table ~\ref{tab:sample}. 

\subsection{Facilities}  
The {\it Swift} satellite \citep{Gehrels2004} and MPG 2.2 m telescope at ESO La Silla, Chile were employed for conducting the observations. All sources were observed in 13 filters; ({\it uvw2, uvm2, uvw1, u, b, v}) of {\it Swift}-UVOT (The Ultraviolet and Optical Telescope, \citealt{Roming2005}) and 7 optical-IR filters ({\it $g^\prime,  r^\prime,  i^\prime, z^\prime, J, H, K_s$ }) of GROND (Gamma-Ray Optical/Near-infrared Detector, \citealt{Greiner2008}). The resulting 13-filter SED covers a wavelength range of 1600 \AA  - 20000 \AA. The main advantage of using the overlap of two filters : $g^\prime$ from GROND and {\it b} from UVOT is to cross-calibrate the two instruments for the combined data analysis.

\subsection{Observing Strategy}
GROND observations were performed as close to the {\it Swift} observations as possible.
While truly simultaneous observations were rarely feasible due to ground visibility constraints, both instruments often observed within 1-2 days of each other. {\it Swift}-UVOT observed in each of its six filters in sequence, while GROND conducted observations in all the 7 filters, simultaneously, which is very crucial for blazars, due to their variable nature. The typical integration times for {\it Swift} were 100 s each in {\it u, b, v} and 200, 240 and 400 s in {\it uvw1, uvm2, uvw2}, respectively, although the exposure times changed based on the brightness of each object. A typical GROND observation had an integration time of $\sim$ 2 min in {\it $g^\prime,  r^\prime,  i^\prime, z^\prime$} and $\sim$ 4.0 min in $J, H, K_s$.


\section{Data Analysis}
\label{sec:data}

\subsection{\it{Swift}-UVOT}
{\it Swift} data were processed through the standard UVOT pipeline procedure \citep{Poole2007a} in order to remove the bad pixels, to flat-field and to correct for the system response. The magnitude extraction per filter was performed with the UVOT task, \texttt{UVOTMAGHIST}. A circular region was selected for the aperture extraction with variable radius in order to maximize the signal to noise ratio. The resulting magnitudes in each filter were corrected for Galactic extinction, utilizing table 5 presented in \citet{Kataoka2008}. The final magnitudes were converted to the AB system and the results are presented in Table~\ref{tab:uvot}. 
\subsection{GROND}
The data reduction procedure for GROND is described in detail in \citet{Kruhler2008}, and here it is only mentioned briefly. The point spread function (PSF) photometric technique was used for {\it $g^\prime, r^\prime, i^\prime, z^\prime $} filters, whereas due to the undersampled PSF in the near infrared, the standard aperture extraction technique was applied for {\it J, H, K$_s$} photometry. The  4 optical filters were calibrated with the stars in the SDSS Data Release 8 \citep{Aihara2011}, which provides the final magnitudes in the AB system.  2MASS stars \citep{Skrutskie2006} were employed for calibration of the near-IR filters. The correction for the  Galactic foreground extinction was performed with measurements in \citet{Schlafly2011b}. The resulting data were converted to the AB system (Table~\ref{tab:grond}).
\subsection{Variability Correction} 
Blazars, in general, exhibit emission which varies on time scale of a few minutes to years. This variability can have a significant impact on the redshift determination via the photometric technique. GROND data are not affected by intrinsic variability as the observations are performed in all seven filters simultaneously. However, since {\it Swift}-UVOT cycles through every filter, in addition to the non-simultaneous GROND-{\it Swift} observations, these variations in the emission could add to the existing uncertainties. \citet{Rau2012} established that the inherent blazar variability introduces a systematic uncertainty of $\Delta$ m = 0.1 mag for each UVOT filter. The other important factor to consider is that the total SED is obtained by combining the UVOT and GROND data, which are two different instruments, which thus need to be calibrated against each other. \citet{Kruhler2011} performed this task by combining {\it Swift}-UVOT and GROND filter curves, utilizing their spectral overlap and established the following calibration relationship:
\begin{equation}
b-g^\prime = 0.15\,(g^\prime-r^\prime)+0.03\,(g^\prime-r^\prime)^2  
\end{equation}

This equation is based on the assumption that a BL Lac SED is represented by a power law and that its slope does not change over the UV-IR regime. This relation which is valid for -1 $\leq (g^\prime-r^\prime) \leq$ 2 was applied to all the UVOT filters before SED fitting.  The UV-Opt-IR SED of BL Lacs is thought to be dominated by non-thermal synchrotron emission. It can be modeled  over the wavelength interval used here as a powerlaw spectrum. Our 13 band photometry covers this energy regime of the SED of a BL Lac, and moreover the absence of any broad lines makes this approximation valid in particular for these kind of blazars. In addition, stellar templates were fitted to these sources to check for non-powerlaw behavior as explained in the next section.

\subsection{SED fitting}
\label{sec:sed_fit}
The \texttt{LePhare v.2.2} program \footnote[1]{http://www.cfht.hawaii.edu/~arnouts/lephare.html} \citep{Arnouts1999a, Ilbert2006} was employed to determine the photometric redshifts for these objects. This program evaluates the difference between the observational and theoretical data based on $\chi^2$ statistic. We selected three separate template libraries to fit the data, independently. Our first library is comprised of 60 power-law SED templates of the form $F_\lambda \propto \lambda^{-\beta} $, such that $\beta$ ranged from 0 to 3 in steps of 0.05, under the assumption that the UV-Optical-NearInfrared regime for BL Lacs can be fit with a single power-law template. In addition, a library of galaxies and galaxy/AGN hybrids \citep{Salvato2009, Salvato2011} as well as a stellar library using templates from \citet{Pickles1998}, \citet{Bohlin1995} and \citet{Chabrier2000} were used. The results from the first two libraries are presented in Table~\ref{tab:result}. None of our SEDs required a contribution from the stellar libraries.

 \begin{figure}[ht!]
 \includegraphics[width=\columnwidth]{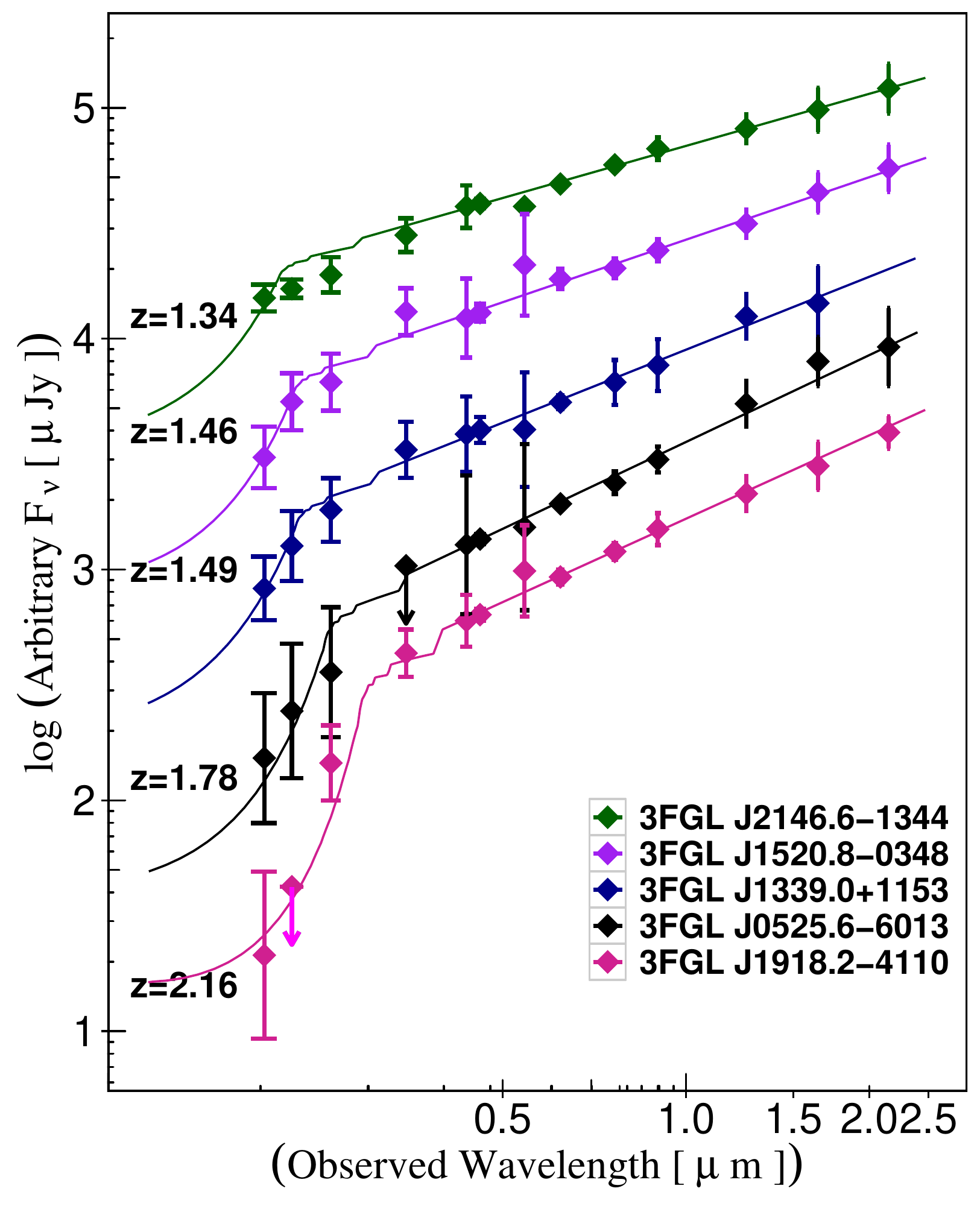}
 \caption{\footnotesize The $Swift$-UVOT + GROND spectral energy distribution for 5 high-$z$ BL Lacs with determined redshifts. The lines show the power law template fits for each source as described in Section \ref{sec:sed_fit}. The details of these fits are provided in Table \ref{tab:result}. The arrows correspond to the upper limits for the measured flux. }
 \label{fig:sed}
 \end{figure}

\section{Results}
\label{sec:results}
The results of SED fitting for 40 sources are presented in Table~\ref{tab:result}. The reliability of our photometric results was determined by Monte Carlo simulations performed in \citet{Rau2012}, where 27000 test SEDs were simulated with $\beta$ ranging from 0.5-2.0 and redshifts 0 to 4. These SEDs were supplied to \texttt{LePhare} to calculate the redshifts and were then compared to the input values. The authors concluded that for sources with simulated redshifts, $z_{sim}$ $>$ 1.2, the photometric redshift reproduced the input value within an accuracy of $\left| \Delta z(1+z_{sim}) \right| < 0.15 $. In addition, a more quantitative selection procedure was applied, which was based on a quantity, P$_{z} = \int f(z) dz$ at $z_{phot} \pm 0.1(1+z_{phot})$, the integral of the probability distribution function. This quantity describes the probability that the redshift of a source is within a factor of 0.1(1+z) of the best fit value. 
Measurements with  P$_{z} > $ 90\% were considered reliable photometric redshifts.
We apply both selection criteria defined above to our SED fitting, which resulted in determining photometric redshifts for 5 sources and establishing upper limits for 20 of them, which is presented in Table~\ref{tab:result}. The $Swift$-UVOT and GROND spectral energy distributions for the new 5 high-$z$ sources, i.e., 3FGL J0525.6-6013, 3FGL J1339.0+1153, 3FGL J1520.8-0348, 3FGL J1918.2-4110, 3FGL J2146.6-1344 is shown are Figure~\ref{fig:sed}.

\section{Discussion}
\begin{figure}[h!]
    \begin{center}
    \includegraphics[width=\columnwidth]{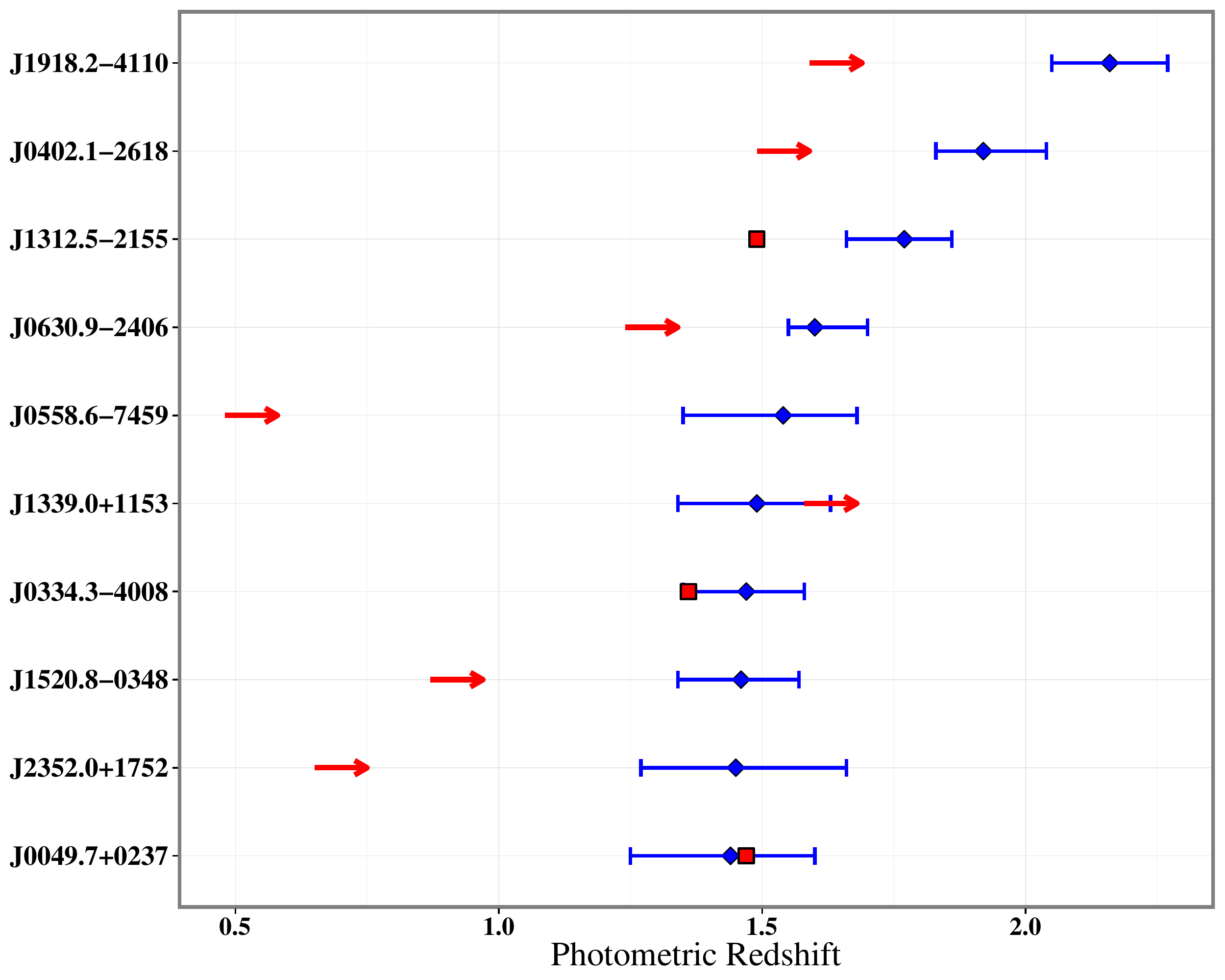}
    \caption{\footnotesize Spectroscopic ($red$ filled squares) vs
      photometric redshifts ($blue$ filled diamonds) comparison for sources  with $z ~>$ 1.3. The red arrows represent the spectroscopic lower limits for these sources. The confirmed spectroscopic redshifts were extracted from 3 LAC data \citep{Ackermann2015} and the lower limits were obtained from \citet{Shaw2013}. The photometric data were obtained from \citet{Rau2012} and this work. The y-axis naming convention follows from the 3FGL catalog.}
    \label{fig:z_comp}
   
     \end{center}
     \end{figure}

 \citet{Ackermann2015} reported 604 BL Lacs out of which 326 sources have redshift measurements of which only 13 with high-$z$ (z $>$ 1.3). 3 of these sources were reported among the 9 high-$z$ BL Lacs by \citet{Rau2012} utilizing the photometric redshift technique. Our study is a continuation of \citet{Rau2012}, who calculate the redshifts for BL Lacs with photometric technique, which increases the sample size by $\sim$ 30\% by finding 5 BL Lacs at $z>1.3$ from our sample of 40 sources. $\sim$ 50\% of the total number of 24 known high-$z$ BL Lacs (5 from this work and 6 from \citet{Rau2012}'s work)  are determined by the photo-$z$ method. A comparison between the spectroscopic and photometric redshift determinations for $z ~>$ 1.3 BL Lacs is presented in Figure~\ref{fig:z_comp}. As seen in this figure, both approaches are consistent with each other. The spectroscopic redshift for the source, 3FGL J1312.5-2155 has been determined by using the absorption features, i.e., C IV and Mg II \citep{Ryabinkov2003}, which could possibly imply a lower limit. This comparison demonstrates that our photometric campaign is an efficient way to uncover rare high-redshift BL Lacs.

\subsection{Blazar Sequence}
 ``Blazar Sequence'' is a scheme, which suggests the existence of a unified model to represent all classes of blazars. Several authors e.g. \citet{Maraschi1995}, \citet{Sambruna1996}, \citet{Fossati1998} proposed this unification idea through observed SEDs for blazars. These authors concluded that the blazar phenomenon is primarily governed by the total luminosity, which is the best indicator for determining the physical properties as well as the radiation mechanisms in these sources. \citet{Fossati1998} noticed several anti-correlations e.g.$~\nu_{sy}^{pk}$ and luminosity at this peak ($L_{sy}^{pk}$), $\nu_{sy}^{pk}$ and the Compton Dominance (CD), and the $\gamma$-ray photon index ($\Gamma_\gamma$). These results indicate that the blazar family lined up on a sequence where more luminous blazars had lower synchrotron peak frequencies, but dominant $\gamma$-ray emission (i.e. CD $>$ 1),  generally the case for powerful FSRQs. On the other hand, the less luminous, high-peaked sources (typically BL Lac objects) have CD $\lesssim$ 1. In other words, blazar sequence predicts the non-existence of high frequency peaked, highly luminous BL Lac objects. A theoretical justification of these correlations are discussed in \citet{Ghisellini1998}.
 
\begin{figure}[ht!]
  
   \includegraphics[width=\columnwidth]{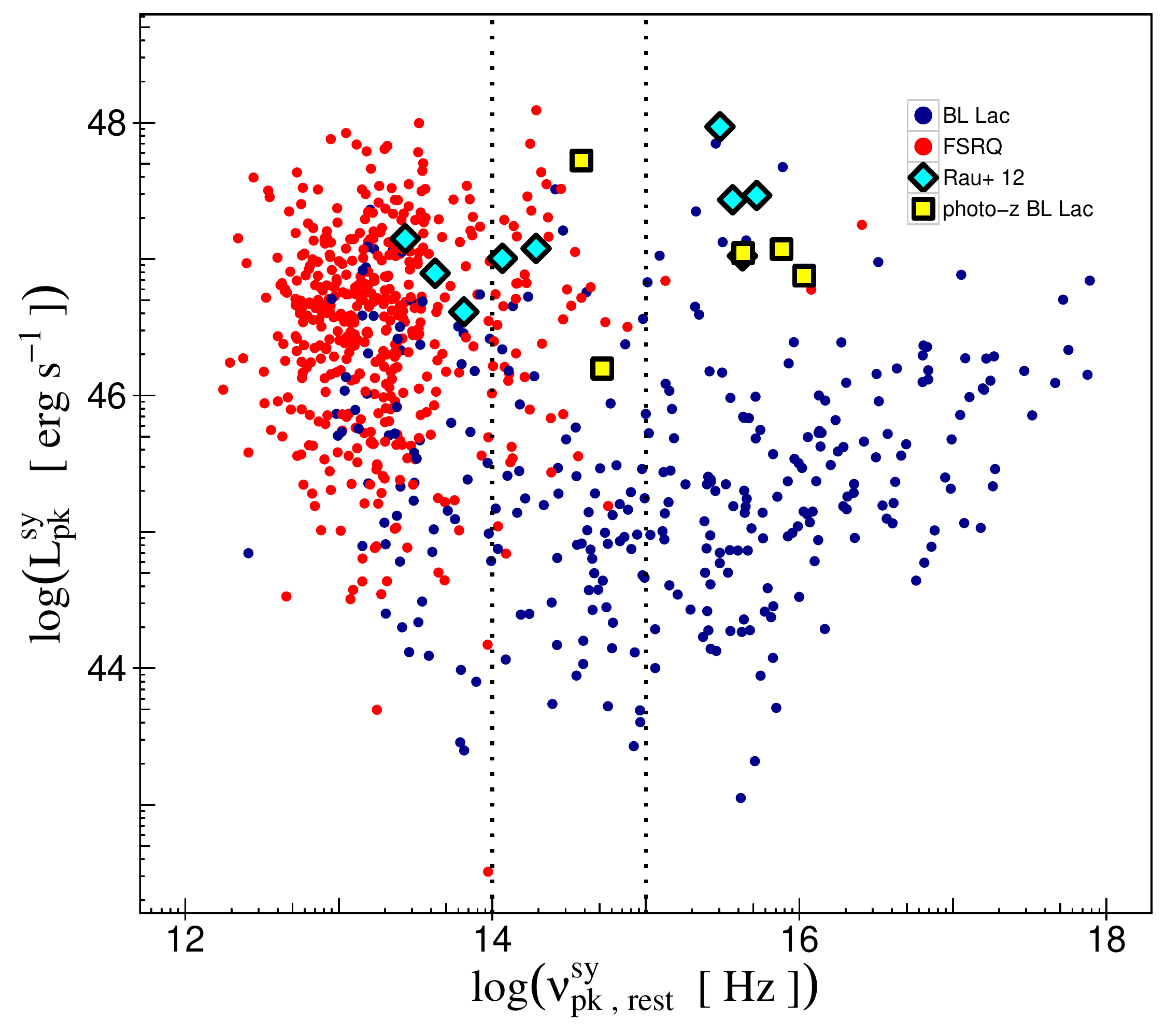}
   \caption{\footnotesize The peak synchrotron frequency in the rest frame ($\nu_{sy}^{pk}$) vs the peak synchrotron luminosity ($L_{sy}^{pk}$). $Yellow$ squares represent 5 new BL Lacs with $z ~>$ 1.3, $cyan$ diamonds are all the BL Lacs from \citet{Rau2012} (including the ones with known redshift via spectroscopic method), $blue$  and $red$ filled circles show  BL Lacs and FSRQs, respectively in the 3LAC data. The vertical dashed lines show the divide between the LSP ($\nu_{sy}^{pk}~<10^{14} \mbox{ [Hz]} $), ISP, ($10^{14}\mbox{ [Hz]}<\nu_{sy}^{pk}~<10^{15}\mbox{ [Hz]}$) and HSP ($\nu_{sy}^{pk}~>10^{15}$) BL Lacs. Note that only blazars with known redshifts are displayed here.} 
   
   \label{fig:vLv}
   \end{figure}

\begin{figure}[h!]
   \includegraphics[width=\columnwidth]{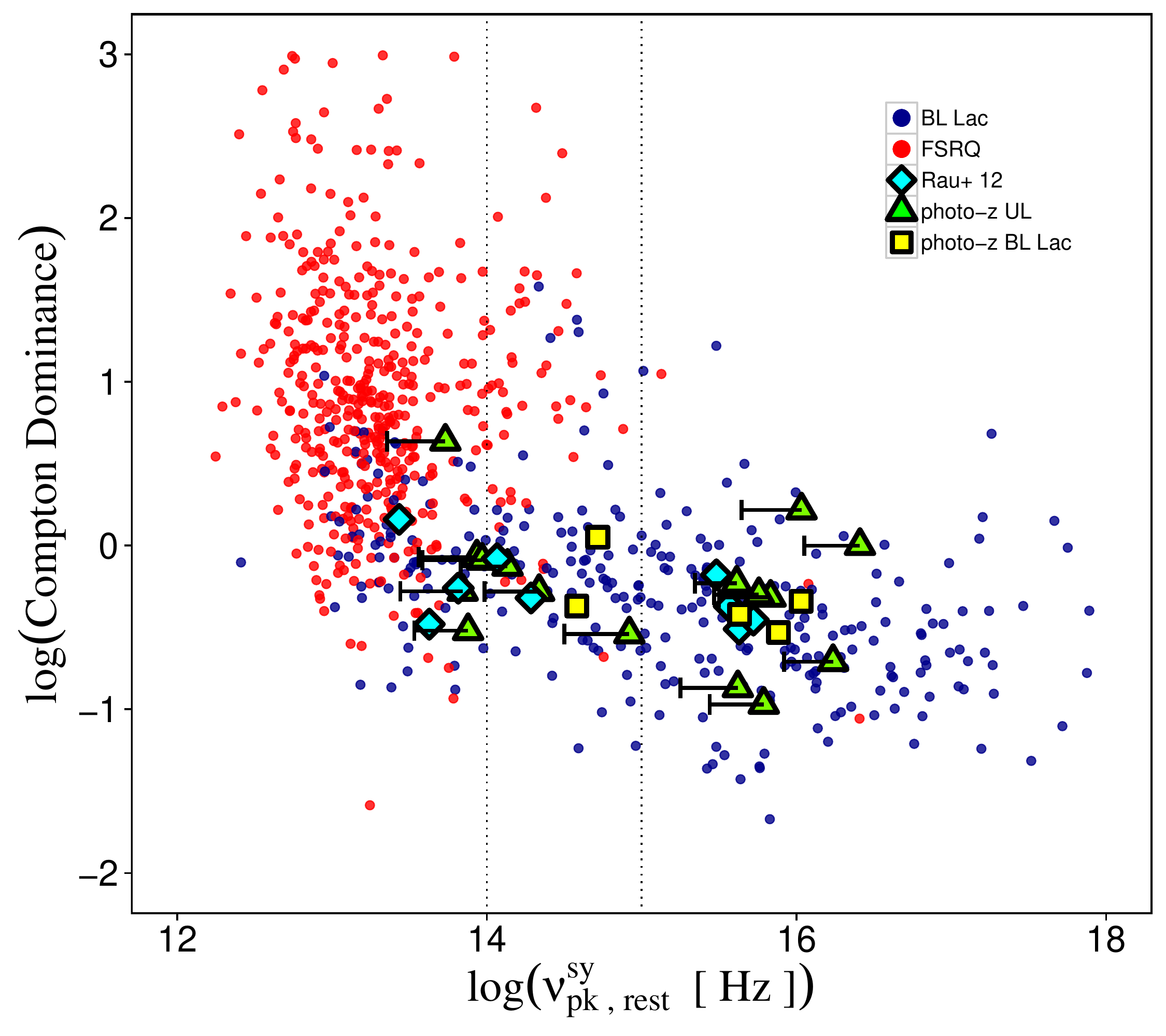}
   \caption{\footnotesize The relationship between the Compton Dominance and  $\nu_{sy}^{\rm pk}$. The color scheme for data points follows from Figure~\ref{fig:vLv}. The $green$ triangles represents 20 sources from our sample with upper limits for photo-$z$s. The `UL` represents the BL Lacs with upper limits for photometric redshifts. The $black$ horizontal bars represent the $\nu_{sy}^{\rm pk}$ with redshift range from zero to the upper limit provided in this work.}
   \label{fig:vcompton}
   \end{figure}

\begin{figure}[h!]
   \includegraphics[width=\columnwidth]{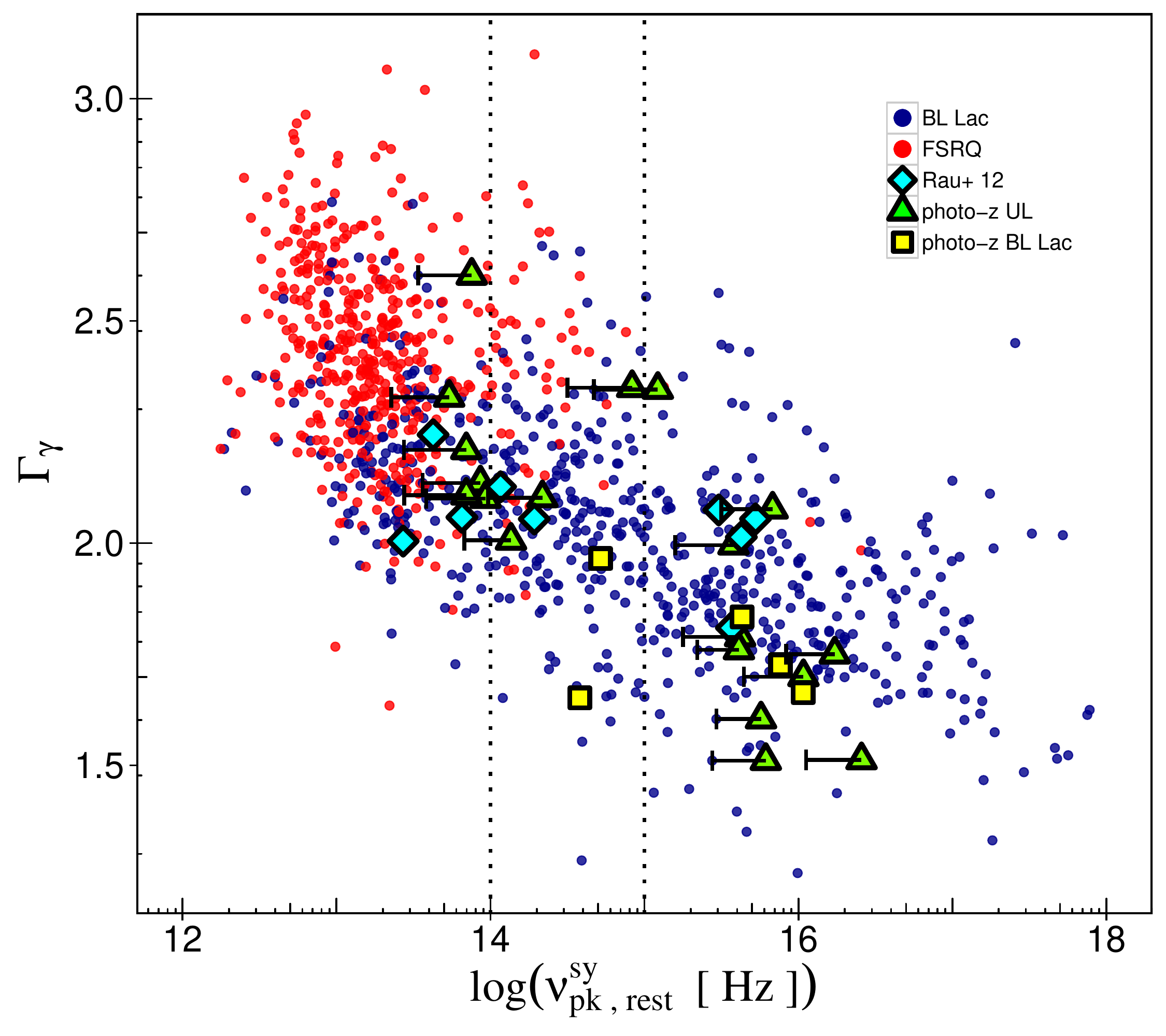}
   \caption{\footnotesize The correlation between the $\gamma$-ray photon index and the redshift corrected peak synchrotron frequency for all the blazars. The $black$ horizontal bars represent the $\nu_{sy}^{\rm pk}$ with redshift range from zero to the upper limit provided in this work. The color scheme for different datasets follows from Figure~\ref{fig:vLv}. Please note that the $\gamma$-ray photon indices have not been corrected for the EBL absorption}. 
   \label{fig:v_gamma}
   \end{figure}

 The concept of a Blazar sequence is an interesting but still debated idea (e.g. \citet{ Padovani2012, Ghisellini2012}). The motivation here is to test how the new high-redshift BL Lacs presented in this study fit in the proposed blazar sequence. We utilize the 3LAC \citep{Ackermann2015} catalog to extract relevant SED parameters, e.g., $\nu_{sy}^{pk}$, $(\nu F_{\nu})_{sy}^{pk}$, Compton Dominance and $\gamma$-ray photon index, for all the blazars present in 3LAC data. $(\nu F_{\nu})_{sy}^{pk}$ was calculated using equation 4 in \citet{Abdo2010} assuming a flat spectrum for blazars in the radio regime,  where the radio flux was obtained from Table 8 in \citet{Ackermann2015}. All the other parameters were extracted from Table 4 in \citet{Ackermann2015}. For those sources in our sample for which CD values were not provided, we calculated them with an online SED fitting tool\footnote{http://tools.asdc.asi.it/SED/}. 
 We calculate luminosities at the peak synchrotron frequencies as they would appear in their rest frame. The obtained parameters are presented in Figure ~\ref{fig:vLv},~\ref{fig:vcompton} and~\ref{fig:v_gamma}. It should be noted that only blazars with known redshifts are plotted in these figures. As can be seen in Figure \ref{fig:vLv}, the five high redshift BL Lac objects (shown with $yellow$ squares) tend to occupy the high $L_{sy}^{pk}$, and a relatively high $\nu_{sy}^{pk}$ region among the known blazar population. We find that all of our high-$z$ BL Lacs are consistent with the known anti-correlation of CD and $\nu_{sy}^{pk}$ (see Figure \ref{fig:vcompton}) with $CD<1$ for all the five high-$z$ BL Lacs, as generally seen in this class of blazars. These findings are in agreement with the theoretical model of \citet{Finke2013} which suggests the existence of sources with high $\nu_{sy}^{pk}$ and high $L_{sy}^{pk}$ but having a CD value less than unity. Furthermore, these objects are hard $\gamma$-ray spectrum sources and we do not find any exception when placed in the $\gamma$-ray photon index vs. $\nu_{sy}^{pk}$ diagram (Figure \ref{fig:v_gamma}). 
 
\subsection{Fermi Blazar Divide}
\citet{Ghisellini2009a} utilized $\sim$100 blazars from the first 3-month survey of {\it Fermi}-LAT data \citep{Abdo2009a} and noticed a division between FSRQs and BL Lacs by comparing their $\gamma$-ray spectral indices ($\Gamma_{\gamma}$) to the $\gamma$-ray luminosities (L$_{\gamma}$). These observations revealed that BL Lacs exhibited harder spectra ($\Gamma_{\gamma}\lesssim$ 2.2) and were less luminous ($L_{\gamma} \lesssim 10^{47} $ erg s$^{-1}$) than FSRQs. It was suggested that this ``Fermi Blazar Divide'' could be interpreted primarily based on the different mass accretion rates with FSRQs having high accretion rates. 
\begin{figure}[hb!]
 \includegraphics[width=\columnwidth]{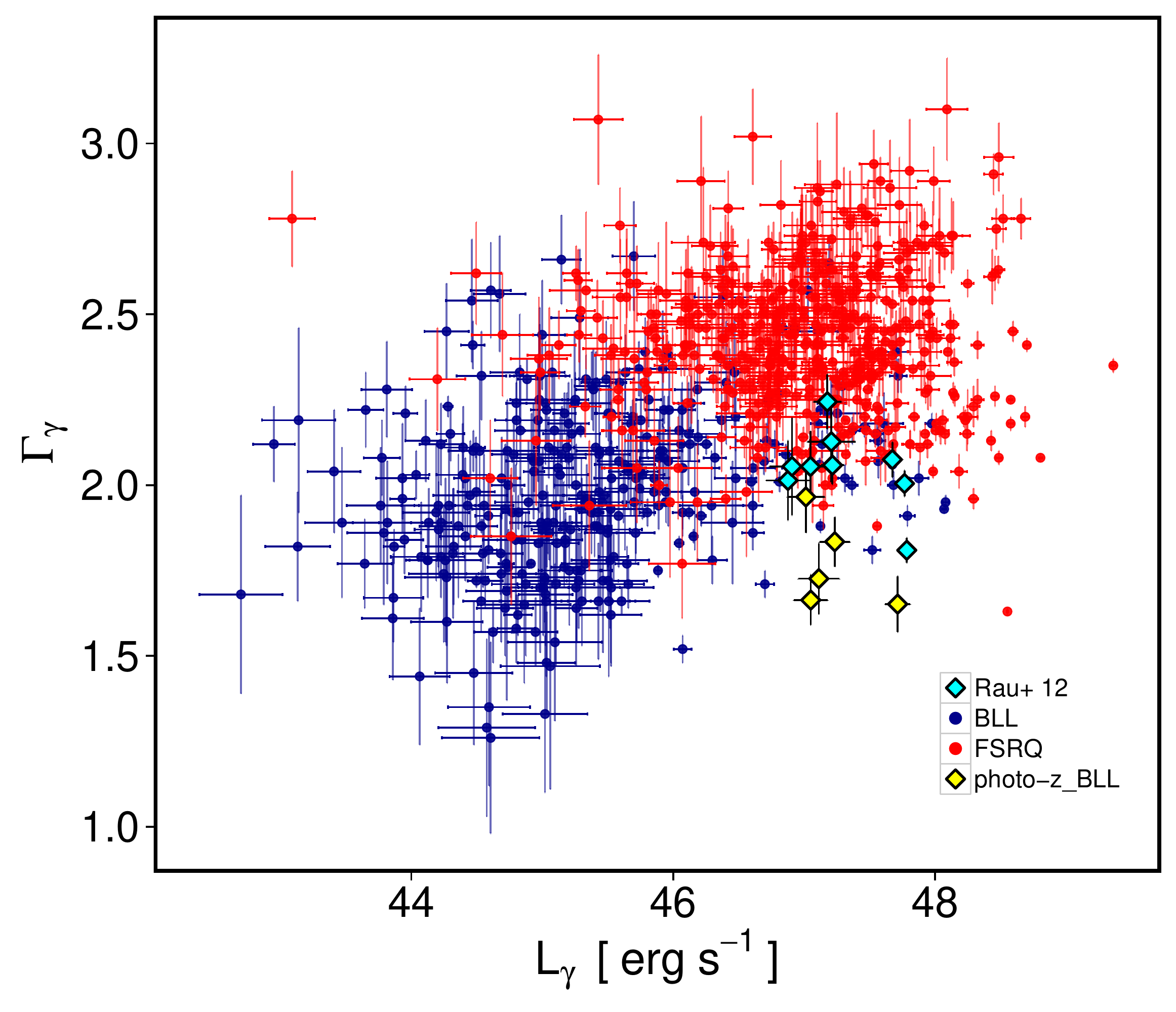}
 \caption{ \footnotesize The correlation between the $\gamma$-ray luminosity (0.1-100 GeV) and the spectral index. The color scheme follows from Figure~\ref{fig:vLv}. Please note that the $\gamma$-ray photon indices have not been corrected for the EBL absorption}.
 \label{fig:gamma_Lgamma}
 \end{figure}

 \begin{figure*}[ht!]
   \begin{center}
   \begin{tabular}{cc}
 \hspace{-1cm}
    \includegraphics[trim=0.3cm 0cm 0cm 0.23cm, clip=true, scale=0.3]{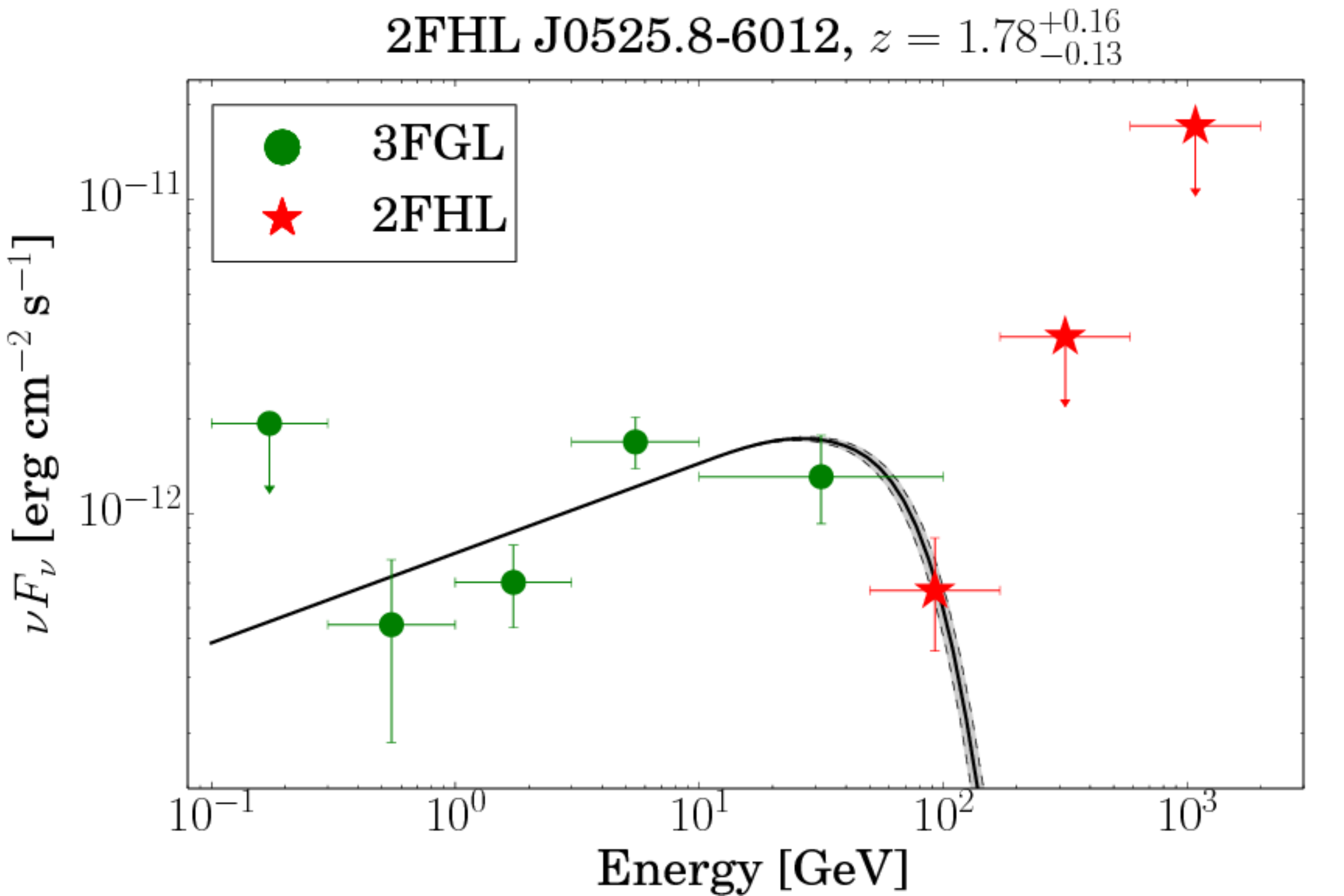}&
   \includegraphics[trim=0.3cm 0cm 0cm 0.23cm, clip=true, scale=0.3]{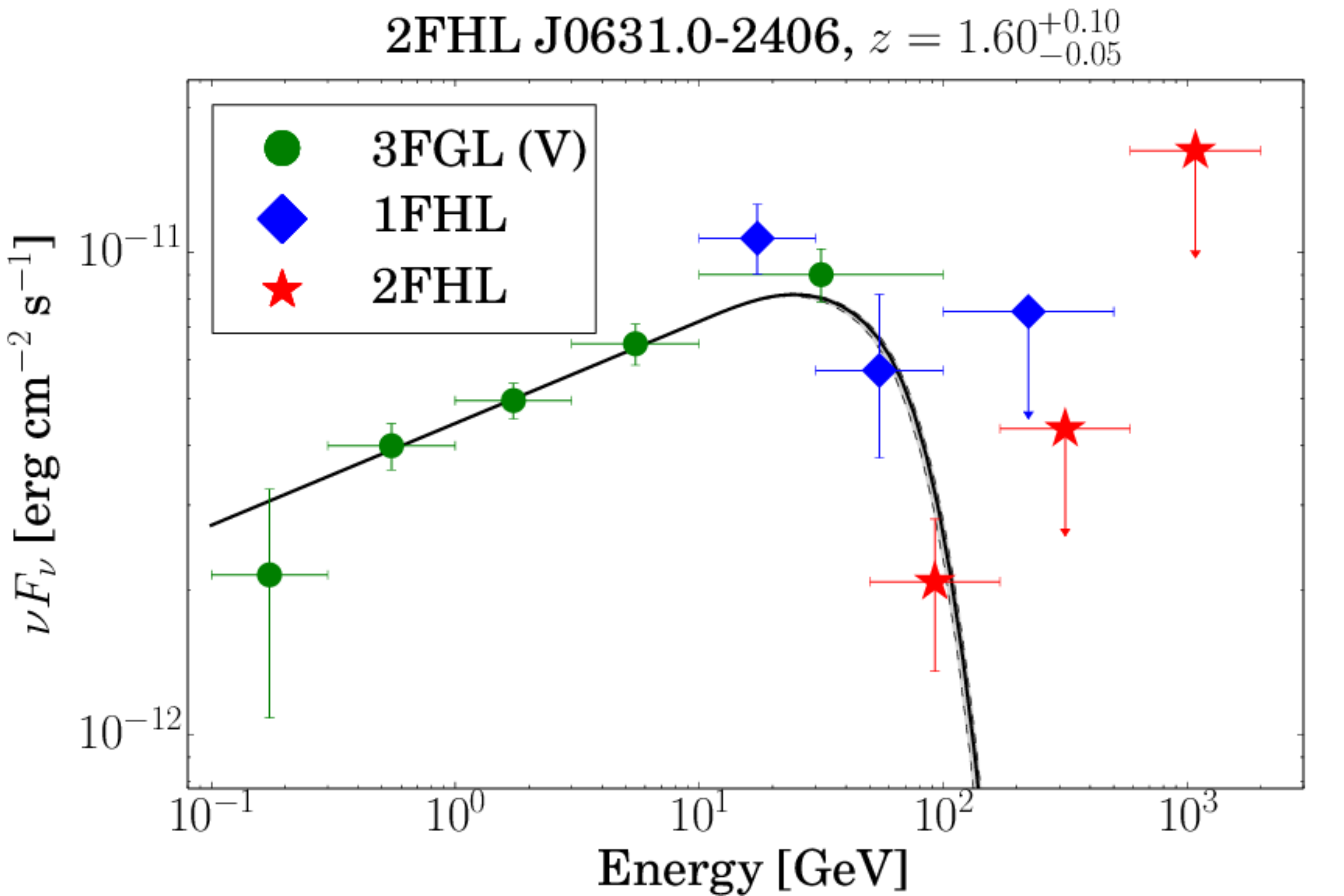}\\
 \hspace{-1cm}
   \includegraphics[trim=0.3cm 0cm 0cm 0.23cm, clip=true, scale=0.3]{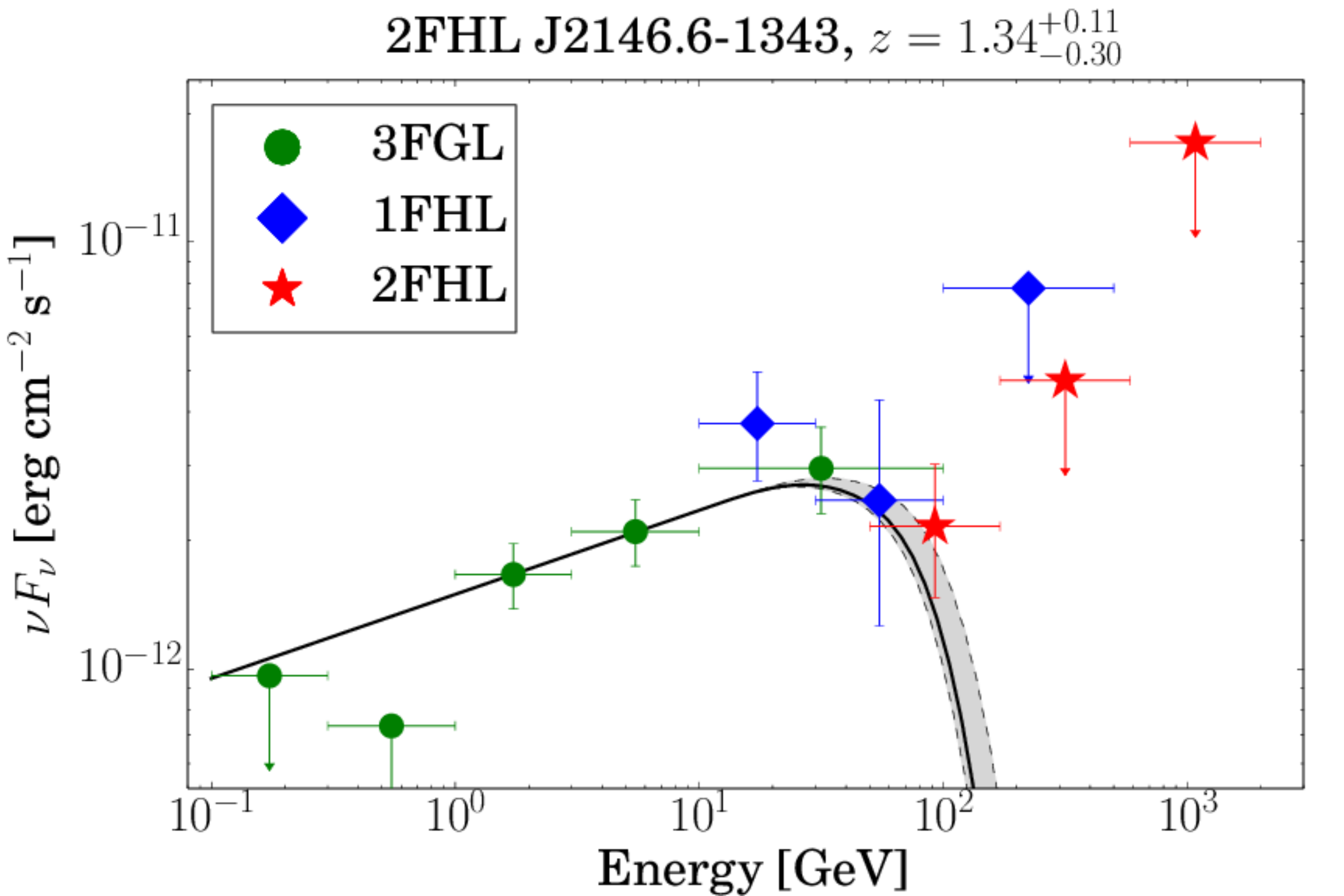}\\
 \end{tabular}
   \end{center}
 \caption{\footnotesize Spectral energy distributions of SUMSS~J052542$-$601341, CRATES~J0630$-$2406, and 1FGL~J2146.6$-$134 obtained combining data reported in the 3FGL, 1FHL and 2FHL catalogs \citep{Acero2015, Ackermann2013a, Ackermann2016}. The black line is a fit to the 3FHL data with a power law that is absorbed by the EBL using the model of \cite{Dominguez2011}. The gray area at high energy denotes the uncertainty in the source attenuation (and correspondingly in the source flux) due to the uncertain redshift.
 \label{fig:2FHL}}
 \end{figure*}
 
 Moreover, this division between two groups of blazars on $\Gamma_{\gamma}$ - $L_{\gamma}$ plane has been discussed by  various authors, e.g., \citet{Padovani2012}, \citet{Ghisellini2012}. The former authors noticed high $L_{\gamma}$ and hard $\Gamma_{\gamma}$ for some of the BL Lacs from \citet{Rau2012}, which was argued against by the latter suggesting that although these BL Lacs fall into the higher end of $\Gamma_{\gamma}$ - $L_{\gamma}$ plane, yet they are consistent with the separation of two populations. In particular, \citet{Ghisellini2012} proposed that the emission region in these sources may lie outside the broad line region (BLR) and therefore the resultant SEDs are similar to BL Lacs. In other words, such objects are `blue' FSRQs with the broad emission lines in their optical spectra are swamped by high level of synchrotron emission. 
 
 In Figure~\ref{fig:gamma_Lgamma}, we show the variation of $L_{\gamma}$ as a function of $\Gamma_{\gamma}$ for 3LAC blazars. In this diagram, we also show high-$z$ BL Lacs obtained in this work and by \citet{Rau2012}.
  The L$_{\gamma}$ for all these sources were calculated using equation 1 in \citet{Ghisellini2009a}. 
  Interestingly, as can be seen in Figure~\ref{fig:gamma_Lgamma}, all the five high-$z$ BL Lacs have a hard $\Gamma_{\gamma}$ ($<2.0$) and their $\gamma$-ray luminosities are $\gtrsim 10^{47}$ erg s$^{-1}$. Now, following \citet{Sbarrato2012}, we have
  \begin{equation}
  L_{\rm BLR} \sim 4L_{\gamma}^{0.93}
  \end{equation}
 where $L_{\rm BLR}$ is the BLR luminosity. 
 This indicates that for $L_{\gamma}\sim1 \times 10^{47}$ erg s$^{-1}$, we have $L_{\rm BLR}\sim2 \times 10^{44}$ erg $^{-1}$. Assuming a fraction 10\% of the disk luminosity is reprocessed by the BLR, the disk luminosity is $\sim 2\times 10^{45}$ erg s$^{-1}$. Since there is no estimate of the central black hole mass available in the literature for the five high-$z$ BL Lacs, we assume an average value of $5\times10^{8}~M_{\odot}$ \citep[e.g.,][]{Sbarrato2012}. 
 
This assumption implies $L_{\rm disk}/L_{\rm Edd}\sim 0.03$, which makes their accretion disk radiatively efficient, similar to what is typically observed in powerful FSRQs. However, these quantities should be considered with some caution due to underlying assumptions \citep[e.g.,][]{Sbarrato2012, Ghisellini2012} and we defer to a future multi-wavelength study for a more detailed discussion. Furthermore, although the $L_{\gamma}$ of these BL Lac objects appears to be similar to FSRQs, yet they are consistent with a separation in the $\gamma$-ray photon index alone, a property that has been established and known since EGRET and it is more robust against the redshifts completeness of the redshift population \citep[e.g.][]{ Hartman1999, Venters2007}.
 \subsection{The Extragalactic Background Light}
 \label{sec:ebl}
\begin{figure}[t!]
  \includegraphics[width=\columnwidth]{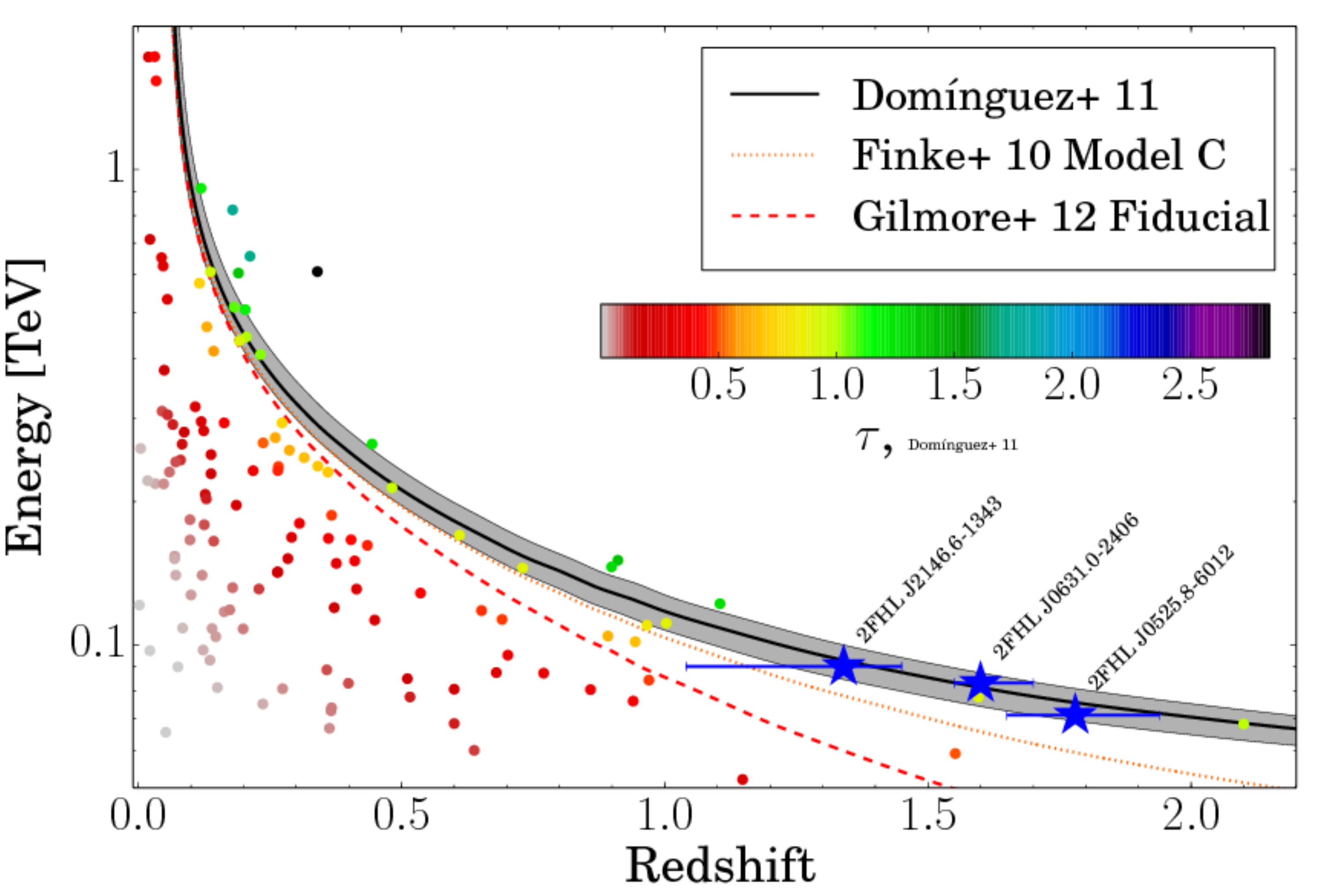}
  \caption{A plot of the highest energy of photons from sources (with E $>$ 50\,GeV) vs. their redshift. The colors of different sources imply their corresponding optical depth ($\tau$) values (see colorbar. Various estimates of the cosmic $\gamma$-ray horizon, obtained from the EBL models by  \citet{Finke2010}(dotted orange line), \citet{Dominguez2011} (solid black line, with its uncertainties as a shaded band) and \citet{Gilmore2011} (dashed red line) are plotted for comparison. The highest energy photons from three of the high-$z$ sources from our sample are consistent with the cosmic $\gamma$-ray horizon ($blue~ filled~stars$) as estimated by \citet{Dominguez2011}.)} 
  \label{fig:energy_z}
  \end{figure}

 Increasing the sample size of high-$z$ BL Lacs is particularly important for probing the EBL , and to better understand the cosmic evolution of the blazar population \citep{Ajello2012}. Photons from sources at $z > $1.3 and energy above 50\,GeV have a probability  $>25$\,\% of interacting with the photons of the UV-optical component of the EBL that leads to the generation of an electron-positron pair.
 For this reason, we looked into which source of the entire photo-$z$ sample \citep[5 sources from this paper and 6 from][] {Rau2012}
 was also reported in the 2FHL catalog of {\it Fermi}-LAT sources detected above 50\,GeV \citep{Ackermann2016}.
 We found three; 1FGL~J2146.6$-$134, CRATES~J0630$-$2406 and SUMSS~J052542$-$601341. Their highest energy photons at 81\,GeV, 83\,GeV, and 71\,GeV, respectively \citep[reported in][]{Ackermann2016} allow us to probe the cosmic $\gamma$-ray horizon (i.e. the distance at which $\tau_{\gamma\gamma}=1$, e.g. \citealt{Dominguez2013}). Figure~\ref{fig:energy_z} shows that those photons are consistent with the $\gamma$-ray horizon in the EBL model developed by \citet{Dominguez2011}. It is important to realize that these high-$z$ sources provide a valuable constraint on the $\gamma$-ray horizon in a region where the sources are scarce.

 Figure~\ref{fig:2FHL} shows the $\gamma$-ray SED of 1FGL~J2146.6$-$134, CRATES~J0630$-$2406 and SUMSS~J052542$-$601341
 obtained combining 3FGL, 1FHL and 2FHL data \citep{Acero2015, Ackermann2013a, Ackermann2016}.
  For each source, we fitted a power law to the 3FGL data and applied to it EBL absorption (in the form  $e^{-\tau(E,z)}$) adopting the \cite{Dominguez2011} model and taking the redshift uncertainty into account. All these sources exhibit hard spectra with photon spectral indices of $\Gamma\approx1.8$, similar to the one of Mrk~421, but belonging to objects that are much farther away. The redshift uncertainty produces a negligible effect in the expected source flux at $z \gtrsim$1.5. As such one may use BL Lacs with photo-$z$ estimates to constrain the EBL in the same way as BL Lacs with spectroscopic redshift have been used in \citet{Ackermann2012}.
 
 Figs.~\ref{fig:energy_z} and \ref{fig:2FHL} show that the $\gamma$-ray horizon is for $z \gtrsim$1.5 is at energy $\lesssim$100\,GeV: i.e. well in the {\it Fermi}-LAT band and well constrained by the first energy bin of the 2FHL catalog.
 For two out of three sources the flux is reduced by a factor $\gtrsim$5 with respect to the unabsorbed flux. 
 
 \section{Conclusions}
 
 This work is a continuation of the photometric redshift determination program utilizing the simultaneous GROND+{\it Swift}-UVOT data described in \citet{Rau2012} which provided photo-$z$ estimates/limits for hundreds of {\it Fermi}. We present 40 sources from the 3FGL catalog, for which redshifts (or upper limits) were determined  by the photometric technique. Five of these sources are high redshift BL Lacs (z$>$1.3). The {\it Fermi} 3LAC catalog comprises of 13 high-$z$ BL Lacs, in addition to the 6 provided by \citet{Rau2012}. By including the 5 BL Lacs in this work, the total number of z $>$ 1.3 sources increases to 24, of which 13 are provided by the spectroscopic method and the other 11 by the photometric technique. Therefore, this work increases the sample size of known high-$z$ BL Lacs by $\sim$30\%.  It should be noted that $\sim$50\% of the known high-$z$ BL Lacs have been found using the photometric redshift method. The latter method is efficient since obtaining data requires considerably shorter integration time than the spectroscopic method as described in Section 2. Moreover, the properties of these objects are in agreement with the blazar sequence which were examined using the various parameters as illustrated in Fig. \ref{fig:vLv}, \ref{fig:vcompton} and \ref{fig:v_gamma}. All the five new sources are classified as LSPs or HSPs exhibiting $L_{sy}^{pk}$ $\simeq$ 10$^{46-48}$ erg s$^{-1}$. The $\gamma$-ray luminosities associated with these objects are $L_{\gamma} \gtrsim 10^{47}$ erg s$^{-1}$ which are higher than the suggested values in the ``Fermi Blazar Divide'' introduced by \citet{Ghisellini2009a} who divided the FSRQs and BL Lacs on $\Gamma_\gamma - L_{\gamma}$ plane, such that the BL Lacs show hard spectra ($\Gamma_{\gamma}$ $<$ 2.2) with low  $\gamma$-ray luminosity ($L_{\gamma} \lesssim 10^{47}$ erg s$^{-1}$). Three of these new objects were reported in the 2FHL catalog, from which the highest energy photons (E $>$ 50 GeV) coupled with the redshift measurements were utilized to test the consistency with the currently known EBL models, e.g. \citet{Finke2010, Dominguez2011, Gilmore2011}. 
Detecting high-$z$ BL Lacs with substantial amount of $>$ 50 GeV emission allowed us to constrain the EBL in a region where there are not, as yet, measurements of the optical depth. The ones derived here show consistency with the prediction of the EBL model of \citet{Dominguez2011}, which is in agreement with the galaxy counts.

\acknowledgments

Part of the funding for GROND (both hardware as well as personnel) was generously granted from the Leibniz-Prize to Prof. G. Hasinger (DFG grant HA 1850/28-1). AK acknowledges funding under NASA contract NNX15AU67G. The authors acknowledge the {\it Swift} team and {\it Swift} PI (N. Gehrels) for scheduling all the {\it Swift}-UVOT observations.


\clearpage
\begin{deluxetable}{lrrrr}
\tablecaption{\label{tab:sample} Target List}
\tablewidth{0pt}
\setlength{\tabcolsep}{0.2in} 
\tablehead{
\colhead{Name}  & \colhead{3FGL Name$^a$} & \colhead{$\alpha_{\rm J2000}^b$} & \colhead{$\delta_{\rm J2000}^b$}  & \colhead{E$_{B-V}^c$}} 
\startdata
CGRaBS J0217+0837       &   J0217.2+0837    & 02 17 02.66 & +08 20 52.4     &   0.13   \\
PMN J0332$-$6155        &   J0331.3$-$6155  & 03 31 18.44 & $-$61 55 28.7   & 0.07 \\ 
CRATES J0505$-$1558     &   J0505.5$-$1558  & 05 05 41.57 & $-$15 58 37.9   & 0.08 \\ 
SUMSS J052542$-$601341  &   J0525.6$-$6013  & 05 25 42.43 & $-$60 13 40.4   & 0.03 \\ 
TXS 0637$-$128 	        &   J0640.0$-$1252  & 06 40 07.01 & $-$12 53 16.1   & 0.39 \\ 
GB6 J0708+2241 	        &   J0708.9+2239    & 07 08 58.31 & +22 41 35.7     & 0.05 \\ 
PMN J0730$-$6602 	&   J0730.5$-$6606  & 07 30 49.42 & $-$66 02 19.3   & 0.14 \\ 
1FGLJ0738.2+1741        &   J0738.1+1741    & 07 38 07.39 & +17 42 18.9     & 0.03 \\
1RXS J084755.9$-$070306 &   J0840.7$-$0651  & 08 47 56.71 & $-$07 03 16.5   & 0.02 \\ 
PMNJ0842-6053           &   J0842.0$-$6055  & 08 42 26.56 & $-$60 53 50.4   & 0.21 \\
1FGL J1059.3$-$1132     &   J1059.2$-$1133  & 10 59 12.60 & $-$11 34 22.1   & 0.03 \\ 
1FGL J1107.8+1502       &   J1107.8+1502    & 11 07 48.05 & +15 02 10.6     & 0.02 \\ 
CLASS J1200+0202     &  J1200.4+0202 & 12 00 52.9 & +12 28 33.2 & 0.02 \\
PMN J1225$-$7313        &   J1225.7$-$7314  & 12 25 35.29 & $-$73 13 39.5   & 0.46 \\ 
1FGL J1226.7$-$1332     &   J1226.9$-$1329  & 12 26 54.41 & $-$13 28 38.9   &	 0.04 \\
FRBA J1254+2211         &   J1254.5+2210    & 12 54 33.27 & +22 11 03.64    &  0.04 \\	
CRATES J1256$-$1146     &   J1256.3$-$1146  & 12 56 15.95 & $-$11 46 37.4   & 0.05 \\ 
SUMSS J132840$-$472748  &   J1328.5$-$4728  & 13 28 40.64 & $-$47 27 49.3   & 0.14 \\
PKS 1326$-$697 	        &   J1330.1$-$7002  & 13 30 11.07 & $-$70 03 12.9   & 0.29 \\  
FRBA J1338+1153	        &   J1339.0+1153    & 13 38 58.99 & +11 53 17.2     & 0.03 \\ 
BZB J1427+2348	        &   J1427.0+2347    & 14 27 00.39 & +23 48 00.0     & 0.05 \\ 
1FGL J1521$-$0350       &   J1520.8$-$0348  & 15 20 49.01 & $-$03 48 51.8   &  0.10 \\ 
CGRaBS J1532$-$1319     &   J1532.7$-$1319  & 15 32 45.37 & $-$13 19 10.1   & 0.12 \\     
CRATES J1552+0850       &   J1552.1+0852    & 15 52 03.26 & +07 50 47.3     & 0.04 \\  
AT20G J1553$-$3118      &   J1553.5$-$3118  & 15 53 33.55 & $-$31 18 30.5   & 0.17 \\ 
BZB J1555+1111 	        &   J1555.7+1111    & 15 55 43.01 & +11 11 24.7       & 0.05 \\ 
BZB J1559+2316	        &   J1559.9+2319    & 15 59 52.20 & +23 16 56.6	    & 0.05 \\ 
CRATES J1610-6649       &   J1610.8-6649    & 16 10 46.46 & $-$66 49 01.3   & 0.09 \\
CGRaBS J1647$-$6438       &   J1647.1-6438    & 16 47 37.74 & $-$64 38 00.27   & 0.17 \\
CRATES J1918$-$4111     &   J1918.2$-$4110  & 19 18 16.01 & $-$41 11 30.8   & 0.09 \\ 
PKS 1942$-$313	        &   J1945.9$-$3115  & 19 45 59.37 & $-$31 11 38.3   & 0.07 \\ 
CRATES J2024$-$4544       &   J2022.2$-$4515  & 20 22 26.26 & $-$45 13 31.5  & 0.03 \\
SUMSS J203451$-$420024  &   J2034.6$-$4202  & 20 34 51.10 & $-$42 00 38.0   & 0.04 \\ 
SDSS J20553$-$00211        &   J2055.2$-$0019  & 20 55 28.23 & $-$00 21 17.2   & 0.09\\	
PMN J2132$-$6624	&   J2131.1$-$6625  & 21 32 37.35 & $-$66 24 27.2   & 0.03 \\ 
CRATESJ2143-3929        &   J2143.1$-$3928  & 21 43 02.82 & $-$39 29 24.0   & 0.02 \\	
1FGL J2146.6$-$1345     &   J2146.6$-$1344  & 21 46 37.01 & $-$13 44 00.9   & 0.04 \\ 
BZB J2158$-$3013        &   J2158.8$-$3013  & 21 58 52.03 & $-$30 13 31.4   & 0.02 \\ 
PKS 2244$-$002          &   J2247.2$-$0004  & 22 47 30.19 & +00 00 06.5    & 0.10 \\
BZB J2334+1408          &   J2334.8+1432    & 23 34 53.83 & +14 32 14.7     &  0.12 \\
\enddata

\tablenotetext{a}{{\it Fermi}/LAT ID.}
\tablenotetext{b}{Coordinates of the optical counterpart with typical uncertainties of 0\farcs3 in both directions.}
\tablenotetext{c}{Reddening extracted from \citet{Schlafly2011b}.}
\end{deluxetable}
\clearpage

\begin{deluxetable}{llllllrrr}
\tablecaption{\label{tab:uvot} {\it Swift}/UVOT photometry}
\tablewidth{0pt}
\tabletypesize{\scriptsize}
\setlength{\tabcolsep}{0.06in} 
\tablehead{
\colhead{Name}  & \colhead{UT Date$^a$} & \multicolumn{6}{c}{AB Magnitude$^b$} &  \colhead{$\Delta  m_{GR-UV}$} \\
\colhead{} & 
\colhead{} & 
\colhead{$uvw2$} & 
\colhead{$uvm2$} & 
\colhead{$uvw1$} & 
\colhead{$u$} & 
\colhead{$b$} & 
\colhead{$v$} & 
\colhead{mag}}
\startdata
CGRaBS J0217+0837 & 2011$-$09$-$13.71 & $ 19.51\pm 0.09 $ & $ 19.54\pm 0.14 $ & $ 18.99\pm 0.10 $ & $ 18.34\pm 0.08 $ & $ 17.89\pm 0.09 $ & $ 16.18\pm 0.10 $  & 1.07 \\  
PMN J0331$-$6155 & 2013$-$06$-$25.13 & $19.36\pm0.10$ & $18.92\pm0.12$ & $18.99\pm0.12$ & $18.94\pm0.16$ & $18.32\pm0.21$ & $18.18\pm0.28$ & $-$0.18 \\
CRATES J0505$-$1558     & 2010$-$12$-$08.45 & $ 19.32\pm 0.18$ & $ 19.29\pm 0.25$ & $ 19.03\pm 0.22$ & $ 18.87\pm 0.26$ & $ 18.37\pm 0.28$ & $ 18.16\pm 0.46$ & $-$0.09\\
SUMSS J052542$-$601341  & 2014$-$02$-$08.62 & $ 22.24\pm 0.28$ & $ 21.73\pm 0.29$ & $ 21.31\pm 0.28$ & $ >20.16 $ & $ 19.93\pm 0.30$ & $ 19.76\pm 0.36$ & 0.71\\
TXS 0637$-$128        & 2014$-$02$-$27.63 & $ 17.31\pm 0.28$ & $ 16.88\pm 0.49$ & $ 17.24\pm 0.28$ & $ 16.87\pm 0.18$ & $ 16.72\pm 0.18$ & $ 16.27\pm 0.32$ & $-$0.01\\
GB6 J0708+2241        & 2014$-$02$-$27.56 & $ 17.85\pm 0.11$ & $ 17.14\pm 0.14$ & $ 17.27\pm 0.14$ & $ 17.33\pm 0.14$ & $ 17.19\pm 0.15$ & $ 16.34\pm 0.21$ & 0.41  \\
PMN J0730$-$6602              & 2014$-$03$-$11.03 & $ 18.50\pm 0.17$ & $ 18.20\pm 0.22$ & $ 18.30\pm 0.24$ & $ 18.13\pm 0.28$ & $ 17.33\pm 0.21$ & $ 16.43\pm 0.18$ & 0.04\\
1FGL J0738.2+1741 & 2012$-$05$-$22.06 & $ 16.88\pm 0.05 $ & $ 16.72\pm 0.08 $ & $ 16.60\pm 0.06 $ & $ 16.25\pm 0.05 $ & $ 15.97\pm 0.06 $ & $ 15.82\pm 0.09 $  & -0.02 \\  
PMN J0842-6053 & 2013$-$03$-$21.45 & $ > 21.38      $ & $ 20.58\pm 0.35 $ & $ 20.24\pm 0.29 $ & $ 19.68\pm 0.26 $ & $ 19.83\pm      $ & $ 18.30\pm 0.33 $  &  0.97 \\  
1RXS J084755.9$-$070306 & 2013$-$06$-$25.47 & $ 18.08\pm 0.07$ & $ 18.00\pm 0.11$ & $ 17.95\pm 0.10$ & $ 17.44\pm 0.11$ & $ 16.97\pm 0.13$ & $ 16.72\pm 0.21$ & $-$0.66 \\
PKS 0944-75 & 2013-04-13.84 & $ > 20.70      $ & $ 19.46\pm 0.29 $ & $ 19.04\pm 0.19 $ & $ 18.70\pm 0.15 $ & $ 18.39\pm 0.16 $ & $ 17.85\pm 0.24 $  & 0.48 \\  
1FGL J1059.3$-$1132     & 2013$-$07$-$25.43 & $ 17.75\pm 0.10$ & $ 17.49\pm 0.11$ & $ 17.19\pm 0.11$ & $ 16.74\pm 0.11$ & $ 16.45\pm 0.14$ & $ 15.93\pm 0.17$ & $-$0.95\\
1FGL J1107.8+1502     & 2013$-$07$-$17.01 & $ 18.40\pm 0.07$ & $ 18.25\pm 0.09$ & $ 18.06\pm 0.09$ & $ 17.67\pm 0.09$ & $ 17.52\pm 0.13$ & $ 17.44\pm 0.25$ & 0.11 \\
CLASS J1200+0202 & 2011$-$04$-$22.09 & $ 24.19\pm 0.24 $ & $ 24.59\pm 0.55 $ & $ 23.85\pm 0.36 $ & $ 23.92\pm 0.86 $ & $ > 21.43      $ & $ > 18.35      $  & 2.34 \\  
PMN J1225$-$7313        & 2013$-$03$-$13.58 &  $>17.78 $ & $>16.59 $ & $ >17.53$ & $ >17.94$ & $ >17.68$ & $ >17.26$ & \nodata \\
1FGL J1226.7-1332 & 2012$-$08$-$13.97 & $ 20.49\pm 0.16 $ & $ 20.06\pm 0.17 $ & $ 20.00\pm 0.22 $ & $ 19.45\pm 0.26 $ & $ > 19.31      $ & $ > 18.68      $  & -0.12\\  
FRBA J1254+2211 & 2011$-$03$-$02.13 & $ 18.47\pm 0.06 $ & $ 18.44\pm 0.09 $ & $ 18.16\pm 0.08 $ & $ 17.77\pm 0.08 $ & $ 17.43\pm 0.09 $ & $ 17.21\pm 0.13 $  & 0.10 \\  
CRATES J1256$-$1146     & 2011$-$05$-$11.11 & $ 18.38\pm 0.05$ & $ 18.18\pm 0.07$ & $ 17.98\pm 0.06$ & $ 17.46\pm 0.06$ & $ 16.58\pm 0.05$ & $ 15.96\pm 0.05$ & 0.49 \\
SUMSS J132840$-$472748  & 2011$-$02$-$26.43 & $ 18.55\pm 0.13$ & $ 18.42\pm 0.22$ & $ 18.22\pm 0.16 $ & $17.85\pm 0.15$ & $ 17.60\pm 0.17$ & $ 17.19\pm 0.21$ & $-$0.23  \\
PKS 1326$-$697        & 2014$-$02$-$17.45 & $ >21.46 $ & $>21.86 $ & $>21.02 $ & $ > 21.01$ & $ 19.02\pm 0.15$ & $ 18.02\pm 0.14$ & 2.75 \\
FRBA J1338+1153       & 2011$-$05$-$11.56 & $ 20.70\pm 0.13 $ & $20.04\pm 0.15 $ & $19.85 \pm8.13$ & $ 19.20\pm 0.12$ & $ 19.03\pm 0.16$ & $ 18.98\pm 0.24$ & 1.08 \\
BZB J1427+2348        & 2014$-$03$-$04.47 & $ 15.18\pm 0.05$ & $ 14.98\pm 0.05$ & $ 14.96\pm 0.05$ & $ 14.45\pm 0.05$ & $ 14.25\pm 0.04$ & $ 13.98\pm 0.04$ & $-$0.03 \\
1FGL J1521$-$0350       & 2014$-$01$-$11.97 & $ 19.18\pm 0.13$ & $ 18.58\pm 0.12$ & $ 18.37\pm 0.12$ & $ 17.60\pm 0.10$ & $ 17.67\pm 0.17$ & $ 17.10\pm 0.22$ & $-$0.50 \\
CGRaBS J1532$-$1319 & 2014$-$04$-$30.21 & $ > 24.40      $ & $ > 23.61      $ & $ > 23.68      $ & $ > 23.08     $ & $ > 22.41      $ & $ > 18.39      $  & 3.25 \\  
CRATES J1552+0850 & 2011$-$09$-$19.36 & $ 20.28\pm 0.26 $ & $ 20.22\pm 0.28 $ & $ 19.54\pm 0.28 $ & $ 19.04\pm 0.29 $ & $ 18.86\pm 0.43 $ & $ 18.52\pm      $  & -0.30 \\ 
AT20G J1553$-$3118      & 2013$-$10$-$03.68 & $ 17.27\pm 0.10$ & $ 17.16\pm 0.13$ & $ 16.96\pm 0.11$ & $ 16.50\pm 0.11$ & $ 16.37\pm 0.12$ & $ 16.07\pm 0.17$ & $-$0.02\\
BZB J1555+1111        & 2014$-$03$-$05.46 & $ 15.33\pm 0.05$ & $ 15.13\pm 0.05$ & $ 15.02\pm 0.05$ & $ 14.60\pm 0.04$ & $ 14.36\pm 0.04$ & $ 14.17\pm 0.05$ & $-$0.13\\
BZB J1559+2316        & 2014$-$03$-$21.07 & $ 19.99\pm 0.12$ & $ 19.79\pm 0.17$ & $ 19.55\pm 0.15$ & $ 18.93\pm 0.14$ & $ 18.51\pm 0.14$ & $ 18.39\pm 0.25$ & 0.11\\
CRATES J1610-6649 & 2013$-$03$-$25.22 & $ 16.20\pm 0.06 $ & $ 15.90\pm 0.06 $ & $ 15.82\pm 0.06 $ & $ 15.39\pm 0.05 $ & $ 15.23\pm 0.05 $ & $ 15.86\pm 0.07 $  & -0.85 \\  
CGRaBS J1647-6438 & 2013$-$04$-$11.98 & $ 18.85\pm 0.17 $ & $ 18.68\pm 0.22 $ & $ 18.12\pm 0.15 $ & $ 17.77\pm 0.15 $ & $ 17.38\pm 0.15 $ & $ 17.36\pm 0.18 $  & -0.56 \\  
CRATES J1918$-$4111     & 2013$-$09$-$26.28 & $ 21.57\pm 0.36$ & $ 20.83\pm $-$1 $ & $19.49 \pm0.16$ & $ 18.30\pm 0.10$ & $ 17.95\pm 0.11$ & $ 17.41\pm 0.19$ & 0.08\\
PKS 1942$-$313        & 2013$-$11$-$23.04 & $ 19.41\pm 0.20$ & $ 19.32\pm 0.28$ & $ 19.02\pm 0.24$ & $ 18.37\pm 0.23$ & $ 17.86\pm 0.26$ & $ 17.47\pm 0.30$ & $-$0.18\\
CRATES J2024-4544 & 2012$-$10$-$23.01 & $ 19.43\pm 0.17 $ & $ 18.91\pm 0.20 $ & $ 18.79\pm 0.19 $ & $ 18.59\pm 0.25 $ & $ > 18.56     $ & $ > 18.45      $  & -0.52 \\  
SUMSS J203451$-$420024  & 2013$-$09$-$29.62 & $ 18.52\pm 0.76$ & $ 18.38\pm 0.09$ & $ 18.45\pm 0.11$ & $ 18.06\pm 0.14$ & $ 17.81\pm 0.01$ & $ 18.10\pm 0.25$ & $-$0.04\\
SDSS J20553-00211 & 2012$-$09$-$29.30 & $ 19.08\pm 0.10 $ & $ 19.33\pm 0.17 $ & $ 19.26\pm 0.17 $ & $ 18.86\pm 0.15 $ & $ 18.50\pm 0.17 $ & $ 18.24\pm 0.27 $  & 0.09 \\ 
 PMN J2132$-$6624              & 2013$-$07$-$02.79 & $ >22.87$ & $>22.15 $ & $>22.12  $ & $>21.47$ & $ >20.76 $ & $>19.72$ & \nodata \\
CRATES J2143-3929 & 2012$-$03$-$19.18 & $ 20.38\pm 0.11 $ & $ 19.89\pm 0.16 $ & $ 19.94\pm 0.21 $ & $ 19.90\pm 0.34 $ & $ 19.33\pm 0.41 $ & $ 18.81\pm 0.56 $  & 0.02 \\  
1FGL J2146.6$-$1345     & 2013$-$04$-$29.97 & $ 17.95\pm 0.05$ & $ 17.85\pm 0.04$ & $ 17.70\pm 0.07$ & $ 17.27\pm 0.07$ & $ 16.96\pm 0.09$ & $ 16.96\pm 0.01$ & $-$0.19\\
BZB J2158$-$3013        & 2014$-$03$-$29.15 & $ 15.34\pm 0.03$ & $ 15.26\pm 0.04$ & $ 15.24\pm 0.04$ & $ 14.90\pm 0.03$ & $ 14.77\pm 0.03$ & $ 14.56\pm 0.03$ & 0.71 \\
PKS 2244-002 & 2013$-$06$-$17.65 & $ 20.29\pm 0.21 $ & $ 19.62\pm 0.21 $ & $ 19.66\pm 0.22 $ & $ 19.09\pm 0.23 $ & $ 18.51\pm 0.24 $ & $ 18.06\pm 0.32 $  & -0.07 \\  
BZB J2334+1408 & 2012$-$10$-$15.07 & $ 20.79\pm 0.24 $ & $ > 20.80      $ & $ 20.43\pm 0.30 $ & $ 20.08\pm 0.32 $ & $ 18.93\pm 0.22 $ & $ > 18.78 $  & 0.22 \\  
\enddata
\tablenotetext{a}{Approximate start time of $uvw2$ exposure.}
\tablenotetext{b}{Corrrected for Galactic foreground reddening. Upper limits are 3\,$\sigma$.}
\tablenotetext{c}{Variability-correction factor to be applied to GROND photometry (see text).} %
\end{deluxetable}
\clearpage
\begin{deluxetable}{lllllllll}
\tablecolumns{11}

\tablecaption{\label{tab:grond}GROND photometry}
\tabletypesize{\scriptsize}
\tablewidth{0pt}
\setlength{\tabcolsep}{0.06in} 
\tablehead{
\colhead{Name}  & \colhead{UT Date$^a$} & \multicolumn{7}{c}{AB Magnitude$^b$} \\
\colhead{} & \colhead{} & \colhead{$g^\prime$} & \colhead{$r^\prime$} & \colhead{$i^\prime$} & \colhead{$z^\prime$} & \colhead{$J$} & \colhead{$H$}  & \colhead{$K_s$}}
\startdata
CGRaBS J0217+0837 & 2011$-$09$-$13.38 & $ 17.40\pm 0.07 $ & $ 15.31\pm 0.16 $ & $  15.26\pm   0.10 $ & $  15.32\pm   0.13 $ & $  15.23\pm   0.05 $ & $  14.86\pm   0.05 $ & $  14.50\pm   0.07 $   \\
PMN J0331$-$6155 & 2013$-$06$-$25.42 & $ 18.22\pm 0.29$ & $ 17.67\pm 0.11$ &  \nodata  &  \nodata & $ 16.57\pm 0.12$ & $ 16.26\pm 0.05$ & $ 16.05\pm 0.05$  \\
CRATES J0505$-$1558     & 2013$-$11$-$13.29 & $ 18.30\pm 0.14$ & $ 17.91\pm 0.20$ & $ 17.07\pm 0.17$ & $ 17.43\pm 0.38$ & $ 16.86\pm 0.07$ & $ 16.42\pm 0.07$ & $ 16.24\pm 0.08$ \\
SUMSS J052542$-$601341  & 2014$-$02$-$09.12 & $ 19.87\pm 0.02$ & $ 19.48\pm 0.01$ & $ 19.26\pm 0.04$ & $ 19.00\pm 0.05$ & $ 18.40\pm 0.09$ & $ 17.94\pm 0.10$ & $ 17.78\pm 0.16$\\
TXS 0637$-$128        & 2014$-$02$-$27.01 & $ 16.67\pm 0.06$ & $ 16.34\pm 0.06$ & $ 16.00\pm 0.10$ & $ 15.83\pm 0.07$ & $ 15.41\pm 0.07$ & $ 15.15\pm 0.10$ & $ 14.97\pm 0.10$\\
GB6 J0708+2241        & 2015$-$01$-$21.10 & $ 17.14\pm 0.04$ & $ 16.81\pm 0.06$ & $ 16.61\pm 0.05$ & $ 16.38\pm 0.08$ & $ 15.75\pm 0.06$ & $ 15.57\pm 0.07$ & $ 15.30\pm 0.07$\\
PMN J0730$-$6602   & 2014$-$03$-$10.99 & $ 17.27\pm 0.03$ & $ 16.89\pm 0.04$ & $ 16.62\pm 0.04$ & $ 16.33\pm 0.05$ & $ 15.93\pm 0.07$ & $ 15.56\pm 0.09$ & $ 15.42\pm 0.08$\\
1FGL J0738.2+1741 & 2012$-$05$-$23.69 & $ 15.91\pm 0.03 $ & $ 15.61\pm 0.03 $ & $  15.35\pm   0.03 $ & $  15.15\pm   0.03 $ & $  14.79\pm   0.05 $ & $  14.38\pm   0.05 $ & $  14.14\pm   0.07 $  \\
PMN J0842$-$6053 & 2012$-$06$-$20.95 & $ 19.79\pm 0.10 $ & $ 19.58\pm 0.06 $ & $  19.26\pm   0.08 $ & $  18.95\pm   0.08 $ & $  18.46\pm   0.08 $ & $  18.05\pm   0.08 $ & $  17.64\pm   0.09 $  \\
1RXS J084755.9$-$070306 & 2014$-$02$-$07.16 & $ 16.92\pm 0.03$ & $ 16.60\pm 0.03$ & $ 16.32\pm 0.05$ & $ 15.98\pm 0.05$ & $ 15.68\pm 0.07$ & $ 15.32\pm 0.08$ & $ 14.93\pm 0.11$\\
1FGL J1059.3$-$1132     & 2014$-$02$-$07.20 & $ 16.39\pm 0.04$ & $ 16.01\pm 0.03$ & $ 15.72\pm 0.04$ & $ 15.51\pm 0.04$ & $ 15.08\pm 0.05$ & $ 14.75\pm 0.08$ & $ 14.37\pm 0.09$\\
1FGL J1107.8+1502     & 2013$-$07$-$16.96 & $ 17.49\pm 0.02$ & $ 17.28\pm 0.03$ & $ 17.85\pm 0.05$ & $ 17.72\pm 0.02$ & $ 16.73\pm 0.07$ & $ 16.46\pm 0.12$ & $ 16.37\pm 0.11$\\
CLASS J1200+0202 & 2011$-$11$-$23.18 & $ 21.30\pm 0.10 $ & $ 20.65\pm 0.06 $ & $  20.54\pm   0.09 $ & $  20.47\pm   0.11 $ & $  19.77\pm   0.17 $ &  \nodata  &  \nodata   \\
PMN J1225$-$7313   & 2013$-$03$-$10.39 & $ 18.31\pm 0.03$ & $ 17.57\pm 0.03$ & $ 17.16\pm 0.05$ & $ 16.92\pm 0.04$ & $ 16.39\pm 0.07$ & $ 16.02\pm 0.08$ & $ 15.83\pm 0.09$ \\
1FGL J1226.7$-$1332 & 2013$-$06$-$24.95 & $ 19.10\pm 0.10 $ & $ 18.10\pm 0.06 $ &  \nodata  & $  17.79\pm   0.10 $ & $  17.34\pm   0.05 $ & $  16.98\pm   0.05 $ & $  16.72\pm   0.07$  \\
FRBA J1254+2211 & 2011$-$03$-$04.38 & $ 17.38\pm 0.03 $ & $ 17.10\pm 0.03 $ & $  16.88\pm   0.03 $ & $  16.68\pm   0.03 $ & $  16.41\pm   0.05 $ & $  16.15\pm   0.05 $ & $  15.80\pm   0.07 $  \\
CRATES J1256$-$1146     & 2014$-$08$-$21.97& $ 16.44\pm 0.03$ & $ 15.62\pm 0.04$ & $ 15.28\pm 0.03$ & $ 14.99\pm 0.02$ & $ 14.63\pm 0.04$ & $ 14.44\pm 0.04$ & $ 14.46\pm 0.08$\\
SUMSS J132840$-$472748  & 2011$-$02$-$28.37& $ 17.58\pm 0.02$ & $ 17.45\pm 0.01$ & $ 17.28\pm 0.03$ & $ 17.09\pm 0.03$ & $ 16.82\pm 0.08$ & $ 16.53\pm 0.15$ & $ 16.45\pm 0.09$\\
PKS 1326$-$697        & 2014$-$02$-$17.35& $ 18.91\pm 0.04$ & $ 18.39\pm 0.02$ & $ 18.22\pm 0.04$ & $ 17.93\pm 0.06$ & $ 17.45\pm 0.08$ & $ 16.86\pm 0.09$ & $ 16.46\pm 0.15$\\
FRBA J1338+1153       & 2014$-$04$-$22.22& $ 18.98\pm 0.05$ & $ 18.68\pm 0.03$ & $ 18.47\pm 0.09$ & $ 18.28\pm 0.11$ & $ 17.75\pm 0.09$ & $ 17.61\pm 0.16$ & $ ...$\\
BZB J1427+2348        & 2014$-$03$-$01.37& $ 14.21\pm 0.06$ & $ 13.97\pm 0.05$ & $ 13.88\pm 0.13$ & $ 13.56\pm 0.03$ & $ 13.31\pm 0.06$ & $ 13.11\pm 0.07$ & $ 12.86\pm 0.06$ \\
1FGL J1521$-$0350       & 2014$-$01$-$12.35& $ 17.61\pm 0.03$ & $ 17.25\pm 0.04$ & $ 17.13\pm 0.04$ & $ 16.94\pm 0.04$ & $ 16.65\pm 0.05$ & $ 16.31\pm 0.08$ & $ 16.05\pm 0.10$\\
CGRaBS J1532$-$1319 & 2014$-$04$-$30.37 & $ 21.94\pm 0.13 $ & $ 19.95\pm 0.04 $ & $  18.74\pm   0.03 $ & $  17.84\pm   0.03 $ & $  16.20\pm   0.05 $ & $  15.22\pm   0.05 $ & $  14.54\pm   0.07 $  \\
CRATES J1552+0850 & 2011$-$09$-$18.98 & $ 18.78\pm 0.03 $ & $ 18.39\pm 0.03 $ & $  18.06\pm   0.03 $ & $  17.82\pm   0.03 $ & $  17.45\pm   0.05 $ & $  17.02\pm   0.05 $ & $  16.74\pm   0.07 $ \\
AT20G J1553$-$3118      & 2013$-$03$-$19.40 & $ 16.35\pm 0.02$ & $ 16.18\pm 0.01$ & $ 15.97\pm 0.04$ & $ 15.88\pm 0.04$ & $ 15.64\pm 0.07$ & $ 15.31\pm 0.11$ & $ 15.09\pm 0.11$\\
BZB J1555+1111      & 2014$-$03$-$05.37 & $ 14.32\pm 0.03$ & $ 14.13\pm 0.03$ & $ 13.94\pm 0.05$ & $ 13.77\pm 0.04$ & $ 13.57\pm 0.07$ & $ 13.30\pm 0.08$ & $ 13.10\pm 0.09$ \\
BZB J1559+2316    & 2014$-$06$-$03.13 & $ 18.44\pm 0.02$ & $ 18.01\pm 0.01$ & $ 17.68\pm 0.03$ & $ 17.37\pm 0.03$ & $ 16.89\pm 0.06$ & $ 16.55\pm 0.09$ & $ 15.92\pm 0.11$ \\
CRATES J1610$-$6649 & 2013$-$03$-$25.34 & $ 15.22\pm 0.03 $ & $ 15.23\pm 0.03 $ & $  14.89\pm   0.03 $ & $  14.90\pm   0.03 $ & $  15.18\pm   0.05 $ & $  14.90\pm   0.05 $ & $  14.90\pm   0.07  $ \\
CGRaBS J1647$-$6438 & 2014$-$04$-$11.98 & $ 17.43\pm 0.04 $ & $ 17.76\pm 0.04 $ & $  17.28\pm   0.04 $ & $  17.14\pm   0.04 $ & $  16.96\pm   0.05 $ & $  16.62\pm   0.05 $ & $  16.36\pm   0.07  $ \\
CRATES J1918$-$4111      & 2013$-$09$-$26.98 & $ 17.88\pm 0.02$ & $ 17.48\pm 0.03$ & $ 17.20\pm 0.03$ & $ 16.96\pm 0.06$ & $ 16.57\pm 0.07$ & $ 16.27\pm 0.10$ & $ 15.91\pm 0.07$ \\
PKS 1942$-$313   & 2013$-$09$-$22.98& $ 17.78\pm 0.02$ & $ 17.29\pm 0.03$ & $ 16.95\pm 0.03$ & $ 16.68\pm 0.06$ & $ 16.17\pm 0.08$ & $ 15.82\pm 0.12$ & $ 15.45\pm 0.10$ \\
CRATES J2024$-$4544 & 2012$-$10$-$23.02 & $ 18.57\pm 0.04 $ & $ 18.62\pm 0.05 $ & $  18.28\pm   0.05 $ & $  18.61\pm   0.05 $ & $  18.76\pm   0.07 $ & $  18.90\pm   0.09 $ & $  18.53\pm   0.10 $ \\
SUMSS J203451$-$420024  & 2013$-$09$-$28.99& $ 17.81\pm 0.04$ & $ 17.76\pm 0.03$ & $ 17.55\pm 0.06$ & $ 17.48\pm 0.03$ & $ 17.28\pm 0.12$ & $ 17.18\pm 0.17$ & $ 16.88\pm 0.10$ \\
SDSS J20553$-$00211 & 2012$-$09$-$29.15 & $ 18.46\pm 0.03 $ & $ 18.25\pm 0.03 $ & $  18.06\pm   0.03 $ & $  17.88\pm   0.03 $ & $  17.65\pm   0.05 $ & $  17.57\pm   0.06 $ & $  17.34\pm   0.07 $  \\
PMN J2132$-$6624  & 2014$-$11$-$21.06& $ 20.68\pm 0.05$ & $ 20.14\pm 0.03$ & $ 19.94\pm 0.06$ & $ 19.56\pm 0.07$ & $ 19.06\pm 0.14$ & $ 19.15\pm 0.24$ & $ 18.40\pm 0.20$ \\
CRATES J2143$-$3929 & 2012$-$03$-$19.40 & $ 19.15\pm 0.05 $ & $ 18.26\pm 0.07 $ & $  17.66\pm   0.08 $ & $  17.75\pm   0.08 $ & $  17.54\pm   0.05 $ & $  17.18\pm   0.06 $ & $  16.89\pm   0.24$  \\
1FGL J2146.6$-$1345     & 2013$-$08$-$26.14 & $ 16.93\pm 0.02$ & $ 16.72\pm 0.01$ & $ 16.51\pm 0.02$ & $ 16.34\pm 0.04$ & $ 16.12\pm 0.06$ & $ 15.92\pm 0.09$ & $ 15.69\pm 0.10$ \\
BZB J2158$-$3013        & 2014$-$11$-$09.12& $ 14.73\pm 0.00$ & $ 14.51\pm 0.03$ & $ 14.37\pm 0.04$ & $ 14.20\pm 0.03$ & $ 13.88\pm 0.07$ & $ 13.75\pm 0.07$ & $ 13.50\pm 0.06$ \\
PKS 2244-002 & 2013$-$06$-$17.43 & $ 18.43\pm 0.03 $ & $ 18.03\pm 0.03 $ &  \nodata  &  \nodata  & $  16.98\pm   0.05 $ & $  16.63\pm   0.05 $ & $  16.20\pm   0.07 $ \\
BZB J2334+1408 & 2012$-$10$-$16.18 & $ 18.86\pm 0.03 $ & $ 18.50\pm 0.03 $ & $  18.27\pm   0.03 $ & $  18.00\pm   0.03 $ & $  17.67\pm   0.05 $ & $  17.46\pm   0.05 $ & $  17.23\pm   0.07$ \\
\enddata
\tablenotetext{a}{Exposure start time.} 
\tablenotetext{b}{Corrected for Galactic foreground reddening. Not corrected for variability. Upper limits are 3\,$\sigma$.}
\end{deluxetable}
\clearpage

\begin{deluxetable}{lll|ccll|lclr}
\tablecolumns{11}
\tabletypesize{\footnotesize}
\tablewidth{0pt}
\tablecaption{\label{tab:result}SED fitting}
\setlength{\tabcolsep}{0.01in} 
\tablehead{
\colhead{Name}  & \colhead{$z_{\rm phot, best}^a$} & \colhead{$z_{\rm spec, img}^b$}  & \multicolumn{4}{c}{power law} & \multicolumn{4}{c}{galaxy} \\
\colhead{} & \colhead{} & \colhead{} & \colhead{$z_{\rm phot}$ $^c$} & \colhead{$\chi^2$} & \colhead{P$_{\rm z}^d$} & \colhead{$\beta^e$} & \colhead{$z_{\rm phot}$ $^c$} & \colhead{$\chi^2$} & \colhead{P$_{\rm z}$  $^d$} & \colhead{model}} 
\startdata
 \multicolumn{11}{c}{\textbf{\small Sources with confirmed photometric redshifts}} \\
\hline
SUMSS J052542-601341  &	${1.78}^{+0.16}_{-0.13}$  & \nodata & ${1.78}^{+0.16}_{-0.13}$ &   10.4 &  98.9 &  1.25 & ${0.02}^{+0.03}_{-0.02}$ & 54.7 &  99.9 & Spi4\_template\_norm.sed \\ 
FRBA J1338+1153     & ${1.49}^{+0.15}_{-0.14}$ & $>$ 1.587  & ${1.49}^{+0.15}_{-0.14}$ &   3.7 &  98.8 &  1.05 & ${1.27}^{+0.06}_{-1.27}$ & 27.2 &  77.5 & I22491\_60\_TQSO1\_40.sed \\  
1FGL J1521.0$-$0350     &	${1.46}^{+0.12}_{-0.11}$  & $>$0.867 & ${1.46}^{+0.12}_{-0.11}$ &   7.6 &  99.7 &  0.90 & ${1.19}^{+0.28}_{-0.05}$ & 27.0 &  81.8 & I22491\_40\_TQSO1\_60.sed \\ 
CRATES J1918$-$4111     &	${2.16}^{+0.11}_{-0.01}$  & $>$ 1.591 & ${2.16}^{+0.11}_{-0.01}$ &   5.6 & 100.0 &  1.20 & ${0.00}^{+0.01}_{-0.00}$ & 128.0& 100.0 & S0\_90\_QSO2\_10.sed \\
1FGL J2146.6$-$1345     &  ${1.34}^{+0.11}_{-0.20}$ & \nodata & ${1.34}^{+0.11}_{-0.20}$ &   15.1 &  93.5 &  0.75 & ${1.45}^{+0.05}_{-0.03}$ & 39.2 & 100.0 & pl\_I22491\_20\_TQSO1\_80.sed   \\ 
\hline \hline\multicolumn{11}{c}{\textbf{\small Sources with photometric redshifts upper limits}} \\
\hline
 CGRaBS J0217+0837  & \nodata & 0.085 & $ { 1.47}_{- 0.06}^{+ 0.01} $ &  153.49 &   82.88 &   1.90 & $ { 0.22}_{- 0.01}^{+0.02} $ &  95.72 &  99.87 &  M82\_template\_norm.sed \\
PMN J0331-6155 & $<1.15$  & \nodata & ${0.03}^{+1.12}_{-0.03}$ &   14.7 &  13.3 &  1.35 & ${0.41}^{+0.22}_{-0.10}$ & 6.9 &   76.3 & Sey18\_template\_norm.sed    \\
CRATES J0505-1558     &	$<1.25$  & \nodata & ${0.02}^{+1.23}_{-0.02}$ &   08.3 &  10.4 &  1.25 & ${0.46}^{+0.13}_{-0.44}$ & 13.5 &  52.9 & I22491\_90\_TQSO1\_10.sed \\  
TXS 0637-128          &	$<1.29$  & \nodata & ${0.02}^{+1.27}_{-0.02}$ &   4.7 &  10.5 &  1.05 & ${0.46}^{+0.46}_{-0.44}$ & 3.6 &   34.5 & I22491\_50\_TQSO1\_50.sed   \\ 
GB6 J0708+2241    &	  \nodata  & \nodata & ${0.03}^{+0.99}_{-0.03}$ &   34.4 &  16.0 &  1.10 & ${0.88}^{+0.08}_{-0.09}$ & 10.6 &  99.7 & I22491\_50\_TQSO1\_50.sed \\ 
PMN J0730-6602        &	$<$1.35  & \nodata & ${1.07}^{+0.28}_{-1.07}$ &   15.4 &  34.9 &  1.20 & ${0.05}^{+0.03}_{-0.04}$ & 18.1 &  99.9 & I22491\_80\_TQSO1\_20.sed \\
 1FGL J0738.2+1741  & $< 1.01$ & 0.424 & $ { 0.38}_{- 0.38}^{+ 0.63} $ &    5.07 &   26.58 &   1.05 & $ { 0.00}_{- 0.00}^{+0.03} $ &  13.38 & 100.00 & I22491\_40\_TQSO1\_60.sed \\
 PMNJ0842$-$6053   & $<1.38$ & \nodata & $ { 1.16}_{- 0.11}^{+ 0.22} $ &   16.63 &   60.72 &   1.35 & $ { 1.17}_{- 0.04}^{+0.08} $ &   8.13 & 100.00 & I22491\_70\_TQSO1\_30.sed \\
 1RXS J084755.9-070306 &	 $<1.23$ & \nodata & ${0.96}^{+0.27}_{-0.96}$ &   9.8 &  34.7 &  1.20 & ${0.02}^{+0.04}_{-0.02}$ & 17.8 &  99.9 & I22491\_70\_TQSO1\_30.sed \\ 
1FGL J1059.3$-$1132     & $<1.37$	  & \nodata & ${1.22}^{+0.15}_{-0.27}$ &   01.9 &  63.6 &  1.20 & ${0.04}^{+0.04}_{-0.04}$ & 28.8 &  99.9 & I22491\_70\_TQSO1\_30.sed    \\        
1FGL J1107.8+1502     &	\nodata  & \nodata & ${1.73}^{+0.10}_{-0.07}$ &  250.4 & 100.0 &  2.15 & ${0.63}^{+0.03}_{-0.03}$ & 175.0 & 99.7 & pl\_QSOH\_template\_norm.sed \\ 
CLASS J1200+0202    & \nodata & \nodata & $ { 3.13}_{- 0.06}^{+ 0.11} $ &   23.53 &   99.36 &   0.75 & $ { 3.03}_{- 0.04}^{+0.02} $ &  22.66 &  95.38 & pl\_QSOH\_template\_norm.sed \\
PMN J1225$-$7313        & \nodata  & \nodata & ${2.27}^{+0.06}_{-2.27}$ &   40.0 &  22.3 &  1.55 & ${0.49}^{+0.06}_{-0.07}$ & 3.7 &   99.7 & I22491\_template\_norm.sed.save \\
1FGL J1226.7$-$1332 & \nodata & 0.456 & $ { 0.00}_{- 0.00}^{+ 0.94} $ &   90.38 &   10.32 &   1.70 & $ { 0.44}_{- 0.01}^{+0.01} $ &  15.69 &  99.97 &  Spi4\_template\_norm.sed \\
FRBA J1254+2211     & $< 1.39$ & 0.386 & $ { 1.23}^{+ 0.16}_{- 0.11} $ &    9.11 &   77.20 &   0.95 & $ { 0.00}^{+ 0.00}_{-0.01} $ &  30.32 & 100.00 & I22491\_50\_TQSO1\_50.sed \\
CRATES J1256$-$1146     &	\nodata  & 0.058 & ${1.28}^{+0.02}_{-0.02}$ &  388.9 & 100.0 &  1.55 & ${0.08}^{+0.02}_{-0.02}$ & 28.5 & 100.0 & M82\_template\_norm.sed \\ 	      
PKS 1326-697          &	\nodata  & \nodata & ${3.06}^{+0.03}_{-0.01}$ &   37.8 & 100.0 &  1.10 & ${2.73}^{+0.05}_{-0.01}$ & 45.1 &  98.8 & Spi4\_template\_norm.sed \\  
SUMSS J132840-472748  &	$<1.44$  & \nodata & ${1.26}^{+0.18}_{-0.33}$ &   21.3 &  65.4 &  0.65 & ${1.53}^{+0.02}_{-0.04}$ & 23.7 &  91.9 & pl\_I22491\_20\_TQSO1\_80.sed \\ 
BZB J1427+2348        &	$<0.86$  & \nodata & ${0.85}^{+0.01}_{-0.85}$ &   09.2 &  29.3 &  0.85 & ${0.05}^{+0.23}_{-0.05}$ & 41.7 &  61.6 & I22491\_40\_TQSO1\_60.sed \\  
CGRaBS J1532$-$1319 & \nodata & \nodata & $ { 4.00}_{- 0.00}^{+ 0.00} $ & 1384.56 &  100.00 &   1.50 & $ { 1.69}_{- 0.01}^{+0.01} $ & 291.08 & 100.00 & S0\_90\_QSO2\_10.sed  \\
CRATESJ1552+0850   & $< 1.54$ & 1.05 & $ { 1.43}_{- 0.18}^{+ 0.11} $ &    4.59 &   66.00 &   1.25 & $ { 0.00}_{- 0.00}^{+0.00} $ &  46.92 &  54.94 & I22491\_70\_TQSO1\_30.sed  \\ 
AT20G J1553$-$3118      &	$<1.23$  & \nodata & ${1.23}^{+0.20}_{-0.26}$ &   10.3 &  75.2 &  0.70 & ${1.50}^{+0.04}_{-0.07}$ & 17.3 &  99.8 & pl\_I22491\_20\_TQSO1\_80.sed \\ 
BZB J1555+1111        &	$<0.95$  & \nodata & ${0.93}^{+0.02}_{-0.33}$ &   10.1 &  42.3 &  0.80 & ${0.24}^{+0.01}_{-0.21}$ & 68.1 &  37.4 & pl\_I22491\_20\_TQSO1\_80.sed \\ 
BZB J1559+2316        &	$<1.43$  & \nodata & ${1.18}^{+0.25}_{-0.43}$ &   9.6 &  56.1 &  1.45 & ${0.03}^{+0.03}_{-0.01}$ & 56.7 & 100.0 & I22491\_template\_norm.sed.save  \\ 
CRATES J1610$-$6649 & \nodata & \nodata & $ { 1.53}_{- 0.07}^{+ 0.02} $ &  129.86 &   99.94 &   0.25 & $ { 1.54}_{- 0.00}^{+0.06} $ & 188.92 &  97.54 & pl\_QSOH\_template\_norm.sed \\
CGRaBS J1647$-$6438 & \nodata & \nodata & $ { 1.33}_{- 0.18}^{+ 0.11} $ &  119.07 &   66.93 &   0.75 & $ { 1.72}_{- 0.02}^{+0.01} $ &  73.04 & 100.00 & pl\_I22491\_30\_TQSO1\_70.sed \\
PKS 1942$-$313          &	 $<1.63$ & \nodata & ${1.38}^{+0.25}_{-0.34}$ &   4.2 &  68.9 &  1.45 & ${0.07}^{+0.02}_{-0.03}$ & 37.7 & 100.0 & I22491\_template\_norm.sed.save   \\ 
CRATESJ2024$-$4544 & $< 1.60$ & \nodata & $ { 1.53}_{- 0.08}^{+ 0.07} $ &   43.35 &   99.75 &   2.05 & $ { 1.10}_{- 0.01}^{+0.02} $ & 152.12 &  99.59 & pl\_QSOH\_template\_norm.sed \\
SUMSS J203451-420024  &	 \nodata  & \nodata & ${1.26}^{+0.20}_{-1.26}$ &   12.7 &  36.0 &  0.55 & ${0.49}^{+0.06}_{-0.07}$ & 18.9 &  66.9 & pl\_TQSO1\_template\_norm.sed  \\ 
SDSSJ20553$-$00211 & $< 1.15$ & 0.407 & $ { 0.96}_{- 0.96}^{+ 0.19} $ &   15.94 &   34.04 &   0.70 & $ { 0.46}_{- 0.01}^{+0.01} $ &  54.89 &  91.46 & pl\_I22491\_20\_TQSO1\_80.sed  \\
PMN J2132-6624        &	 \nodata & \nodata & ${2.63}^{+0.32}_{-0.09}$ &   6.1 &  96.6 &  1.25 & ${0.00}^{+0.04}_{-0.00}$ & 14.0 &  99.9 & M82\_template\_norm.sed \\ 
CRATESJ2143$-$3929 & \nodata & 0.429 & $ { 0.93}_{- 0.93}^{+ 0.26} $ &   64.78 &   31.60 &   1.40 & $ { 0.31}_{- 0.03}^{+0.03} $ &  20.84 &  99.60 &  Spi4\_template\_norm.sed \\
 BZB J2158$-$3013       &	$<1.07$ & 0.116 & ${0.03}^{+1.04}_{-0.03}$ &   04.1 &  14.4 &  0.70 & ${0.87}^{+0.04}_{-0.09}$ & 21.2 &  96.5 & pl\_I22491\_20\_TQSO1\_80.sed   \\
PKS2244$-$002      & $< 1.53$ & 0.949 & $ { 1.40}_{- 0.11}^{+ 0.13} $ &    5.42 &   92.50 &   1.35 & $ { 1.29}_{- 0.04}^{+0.03} $ &  67.49 &  96.66 & I22491\_70\_TQSO1\_30.sed  \\
BZBJ2334+1408      & \nodata & 0.415 & $ { 1.71}_{- 0.05}^{+ 0.12} $ &  133.84 &   92.01 &   1.35 & $ { 0.04}_{- 0.00}^{+0.00} $ & 142.54 & 100.00 &  M82\_template\_norm.sed \\
\enddata
\tablenotetext{a}{Best photometric redshift.}
\tablenotetext{b}{Spectroscopic redshift, if known.}
\tablenotetext{c}{Photometric redshifts with 2$\sigma$ confidence level}
\tablenotetext{d}{Redshift probability density at $z_{phot} \pm 0.1(1+z_{phot})$}
\tablenotetext{e}{Spectral slope for power law model of the form $F_{\lambda} \propto \lambda^{-\beta}$}
\end{deluxetable}
\clearpage

\bibliographystyle{apj}
\bibliography{bibliography}
\end{document}